\documentclass[sigconf,review]{acmart}
\usepackage{listings}
\usepackage{xcolor}

\definecolor{codegreen}{rgb}{0,0.6,0}
\definecolor{codeblack}{rgb}{0,0,0}
\definecolor{codegray}{rgb}{0.5,0.5,0.5}
\definecolor{codepurple}{rgb}{0.58,0,0.82}
\definecolor{backcolour}{rgb}{0.95,0.95,0.92}

\lstdefinestyle{mystyle}{
  keywordstyle=\color{magenta},
  numberstyle=\tiny\color{codegray},
  stringstyle=\color{codepurple},
  basicstyle=\ttfamily\footnotesize,
  breakatwhitespace=false,         
  breaklines=true,                 
  captionpos=b,                    
  keepspaces=true,                 
  numbersep=5pt,                  
  showspaces=false,                
  showstringspaces=false,
  showtabs=false,                  
  tabsize=2
}

\AtBeginDocument{%
  }

\renewcommand\footnotetextcopyrightpermission[1]{} 
\usepackage{caption}
\usepackage{subcaption}
\usepackage{tabularx}
\begin{document}

\title{AtomXR: Streamlined XR Prototyping with Natural Language and Immersive Physical Interaction}


\author{Alice Cai}
\email{acai@college.harvard.edu}
\affiliation{%
  \institution{Harvard University}
  \streetaddress{150 Western Ave}
  \city{Cambridge}
  \state{Massachusetts}
  \country{USA}
  \postcode{02139}
}
\author{Caine Ardayfio}
\email{cardayfio@college.harvard.edu}
\affiliation{%
  \institution{Harvard University}
  \streetaddress{150 Western Ave}
  \city{Cambridge}
  \state{Massachusetts}
  \country{USA}
  \postcode{02139}
}
\author{AnhPhu Nguyen}
\email{anugyen1@college.harvard.edu}
\affiliation{%
  \institution{Harvard University}
  \streetaddress{150 Western Ave}
  \city{Cambridge}
  \state{Massachusetts}
  \country{USA}
  \postcode{02139}
}
\author{Tica Lin}
\email{mlin@g.harvard.edu}
\affiliation{%
  \institution{Harvard University}
  \streetaddress{150 Western Ave}
  \city{Cambridge}
  \state{Massachusetts}
  \country{USA}
  \postcode{02139}
}
\author{Elena Glassman}
\email{eglassman@g.harvard.edu}
\affiliation{%
  \institution{Harvard University}
  \streetaddress{150 Western Ave}
  \city{Cambridge}
  \state{Massachusetts}
  \country{USA}
  \postcode{02139}
}
\renewcommand{\shortauthors}{Cai et al.}



\definecolor{codegreen}{rgb}{0,0.4,0}
\definecolor{codeyellow}{rgb}{0.8,0.8,0}
\definecolor{codeblue}{rgb}{0,0,1}
\definecolor{codeorange}{rgb}{.6, .4, 0}

\lstdefinestyle{custom}{
    basicstyle=\ttfamily\small,
    breaklines=true,
    language=Java,
    numbers=left,
    showstringspaces=false,
    tabsize=4,
    keywordstyle=\color{codepurple},
    numberstyle=\color{codegray},
    stringstyle=\color{codeorange},
    identifierstyle=\color{codegreen},
    columns=fullflexible
}
\lstset{style=custom}

\begin{abstract}
As technological advancements in extended reality (XR) amplify the demand for more XR content, traditional development processes face several challenges: 1) a steep learning curve for inexperienced developers, 2) a disconnect between 2D development environments and 3D user experiences inside headsets, and 3) slow iteration cycles due to context switching between development and testing environments. To address these challenges, we introduce AtomXR, a streamlined, immersive, no-code XR prototyping tool designed to empower both experienced and inexperienced developers in creating applications using natural language, eye-gaze, and touch interactions. AtomXR consists of: 1) AtomScript, a high-level human-interpretable scripting language for rapid prototyping, 2) a natural language interface that integrates LLMs and multimodal inputs for AtomScript generation, and 3) an immersive in-headset authoring environment. Empirical evaluation through two user studies offers insights into natural language-based and immersive prototyping, and shows AtomXR provides significant improvements in speed and user experience compared to traditional systems.

\end{abstract}



\received{14 Sep 2023}
\received[revised]{14 Sep 2023}
\received[accepted]{14 Sep 2023}

\maketitle
\pagestyle{plain}

\begin{figure*}
    \centering
  \includegraphics[width=0.8\textwidth]{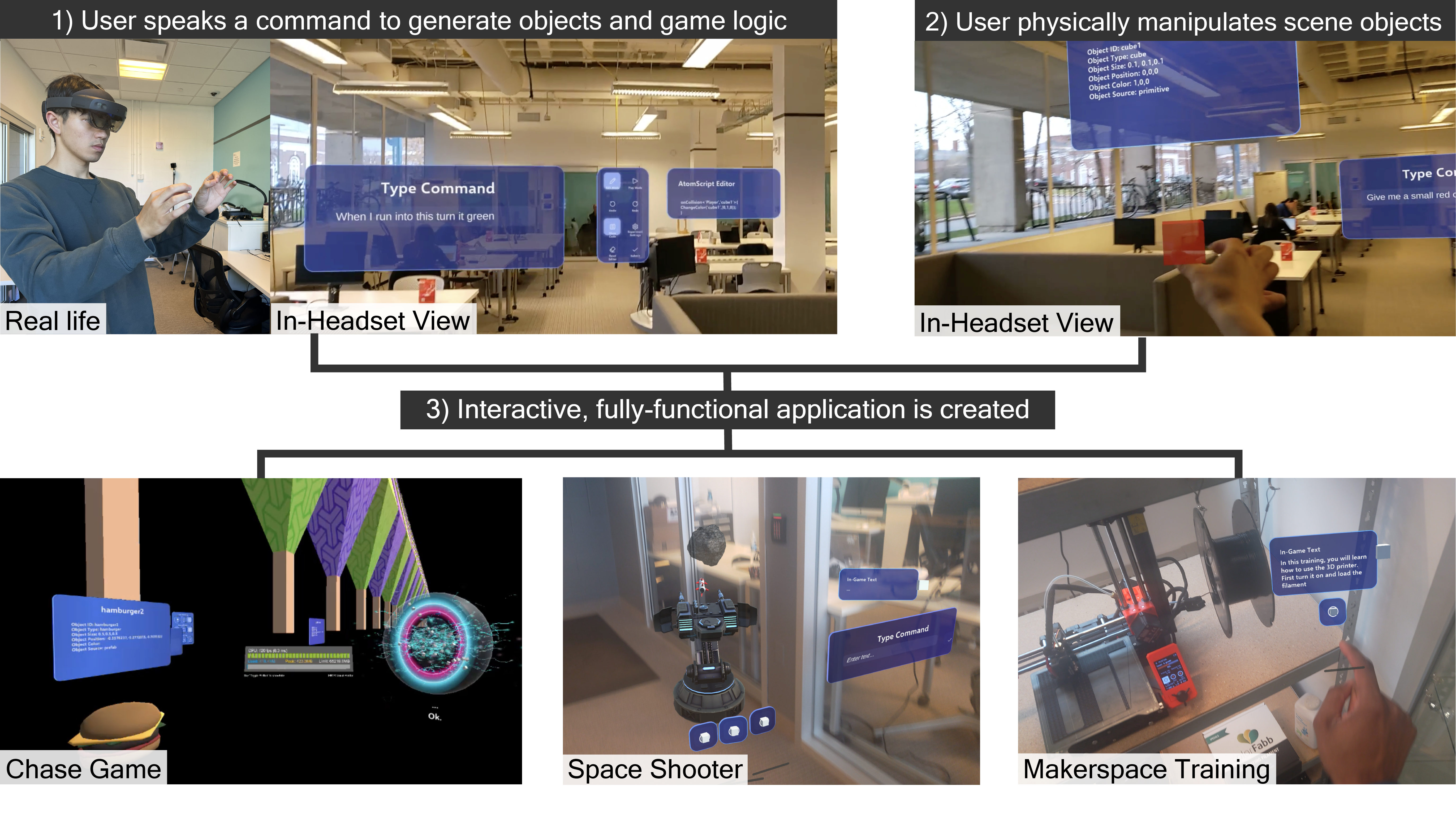}
    \caption{System Preview of AtomXR: In the top left image, a user speaks commands to make objects and game logic. The top-right image features an in-headset view of a user using gestures to manipulate the scene. The bottom row shows several games made with AtomXR, including the chase game, where a user must run away from a chaser. The middle game is the space shooter, where the user moves the turret left and right to shoot at incoming asteroids. The bottom right image is a makerspace training experience teaching user's how to use the equipment in a makerspace. Each of these 3 experiences was created by speaking natural language commands describing the experience in AtomXR.}
    \Description{The figure shows the system preview of AtomXR. In the top left image it shows a man with two hands out making a gesture. A user speaks commands to make objects and game logic. The top-right image features an in-headset view of a user using gestures to manipulate the scene. The bottom row shows several games made with AtomXR, including the chase game, where a user must run away from a chaser. The middle game is the space shooter, where the user moves the turret left and right to shoot at incoming asteroids. The bottom right image is a makerspace training experience teaching user's how to use the equipment in a makerspace. Each of these 3 experiences was created by speaking natural language commands describing the experience in AtomXR.}
\end{figure*}

\section{Introduction}

The recent revival of the extended reality (XR) industry, propelled by the pandemic-driven shift toward virtual activities as well as technical advancements in computing and display hardware, calls for easy and effective XR application development tools. XR bears potential to impact a wide range of fields, spanning education, entertainment, productivity, training, manufacturing, and more \cite{cardenas2022extended}. Yet existing XR development processes produce significant friction due to their inefficiency and substantial barriers to entry. These barriers include the steep learning curve of development tools, the mismatch between development environments on 2D screens, and the 3D user experience inside the headset, and the long testing and iteration cycles necessary in conventional workflows where applications are compiled on the computer and then deployed onto the target headset. Not only do these inefficiencies of XR development practices slow progress in the field, they also exclude those without technical or programming experience, reducing the diversity, richness, and potential utility of applications created in XR. 

To address these issues, we introduce a streamlined application development system, AtomXR, that uses hybrid interaction methods (voice, eye gaze, touch) to enable no-code immersive authoring. While no-code development tools have been desired in this space for a long time, only recently have advancements in natural language processing (NLP), such as the development of large language models (LLMs) like GPT-3~\cite{gpt3} and GPT-4~\cite{koubaa2023gpt}, enabled this kind of system. AtomXR aims to both improve the prototyping process for experienced developers and empower non-experts to create XR applications.

Specifically, AtomXR allows users to create, update, and delete objects and application logic without code. Instead, these operations are done using natural language, eye-gaze, and touch interactions while inside the headset. Users can describe what they want to build, and our system translates their natural language requests into AtomScript, a high-level interpreted language that aims to be easy to read by users and easy to generate using language models. Using eye-gaze and gesture interactions, users can then reference and manipulate objects in the scene to further edit and build the application. In enabling rapid, low-code development of complex experiences, we aim to drive progress towards a new era of AI-enabled and accessible immersive authoring.

We investigated the usefulness and interaction dynamics of the AtomXR system through two experiments involving a total of 24 participants with various levels of prior technical and XR experience. The first experiment compared user performance on a series of development tasks in AtomXR with the traditional Unity-based desktop system. Our system demonstrated significant improvements in accessibility, efficiency, usability, and new feature discovery. Participants were able to complete tasks 2-5 times as fast in the AtomXR systems, and were able to complete more than 2 times as many tasks. Additionally, AtomXR was perceived to be significantly easier to learn and more intuitive to use. The second experiment investigated user interaction dynamics with the major components of the AtomXR system, including programming abstraction, natural language-driven logic generation, and immersive authoring. We synthesize experimental insights on these features and discuss future usage and implications.

Our contributions include both engineering efforts and experimental insights, and are summarized as follows:

\begin{itemize}
\item AtomXR, a no-code, XR development engine supported by:
\begin{itemize}
\item A multimodal approach to interpreting user intent using speech, physical interactions, and eye gaze.
\item AtomScript, a high-level programming language for XR applications.
\end{itemize}
\item Two studies evaluating the effectiveness and user experience of 1) AtomXR compared to a traditional Unity-based system and 2) individual features provided by AtomXR.
\end{itemize}

\section{Related Work}


\subsection{3D \& Immersive Authoring}

While individual 3D development systems vary in the processes they use, there are common capabilities shared among nearly all development systems. Generally, these systems support creating, deleting, and updating the properties of virtual objects, arranging them in a scene, and defining their interactions and behaviors over time. Popular game engines, such as Unity \cite{unity} and Unreal Engine \cite{unreal} are based on the entity-component architecture, which allows developers to easily add and modify an object's properties and behaviors by adding, modifying, or deleting components associated with the object. 
Other approaches to 3D development include WebXR methods, which use libraries and frameworks like Three.js \cite{threejs} and A-FRAME \cite{aframe} that enable browser-based creation and rendering of 3D applications. All these development environments, however, suffer from steep learning curves when adding functional components as they require prior knowledge of programming languages like C\# and Javascript.

A number of development tools have emerged to reduce friction and complexity in XR development. DART \cite{macintyre2004dart} was a collection of extensions to a multimedia programming environment that enabled modular composition of functionality for rapid AR prototyping. ProtoAR \cite{nebeling2018protoar} generates mobile AR views from 2D drawings on paper and Playdoh, but only supports object creation/management and does not support logic generation. Earlier authoring tools like VRML97 \cite{nadeau1999building} and X3D \cite{brutzman2010x3d} use event-passing mechanisms to define user interactions. More recent examples include Unreal Engine’s Blueprints \cite{sewell2015blueprints}, a node-based visual scripting system, and Alice \cite{pausch1995alice}, a block-based programming environment that allows users to rapidly prototype 3D animations. 

Because these development systems are 2D screen-based, the mismatch between the development and testing environment leads to context switching, loss of spatial information, and long iteration cycles. Immersive authoring in XR ~\cite{lee2005immersive} enables developers to more accurately gauge the end user experience inside the headset while developing without context switching. Prior work, drawn from the~\cite{flowmatic} survey of literature, has explored immersive world building and modeling tools like Tilt Brush~\cite{tiltbrush}, Microsoft's Maquette~\cite{maquette}, ScultUp~\cite{sculptup}, and many more~\cite{3dm, isaac, vrwork, liftoff,  medium}. However, robust logic generation is still challenging with these tools. Adding experience logic usually requires text-based scripting, which is slow and unreliable inside head-mounted display (HMD) environments due to the lack of efficient text input methods. One solution to this challenge is the use of visual programming languages, which allow users to create logic and control flow using visual elements rather than text-based scripts.

In one of the earliest immersive visual scripting systems, Steed et al.~\cite{steed1996dataflow} proposed the use of visual dataflow to define object behavior in headset authoring environments, wherein users could draw wires between objects to pass data between them. 
Flowmatic~\cite{flowmatic} presents a drag-and-drop visual scripting interface in VR that allows users to create reactive behaviors while inside the headset.
More recently, companies developing XR platforms like Meta and Microsoft have also begun developing their own immersive authoring environments, such as BuilderBot~\cite{builderbot} and Frame\cite{framevr}. 

However, existing immersive authoring systems often lack expressive power and suffer from complex visual interfaces that lack natural interaction methods. In contrast to existing authoring systems, AtomXR leverages natural language interaction in combination with natural physical interactions for immersive authoring, which allows users to directly and flexibly describe their target application and minimizes the need for users to map intention to text-based or visual scripting.


\subsection{Low-Code Application Development}

In recent years, a growing trend has emerged in favor of no-code and low-code platforms enabling users to build applications without requiring programming skills~\cite{LowCode}. These platforms use various methods of program  synthesis~\cite{ProgramSynthesis} and code generation to enable users to build and customize their applications using non-programming methods like natural language descriptions. This has the potential to greatly expand the number of people who are able to develop applications, increasing the diversity and creativity of the content available.

Low-code development for visual and 3D applications often employ template-based methods, in which users can procedurally generate terrains and even simple runtime behaviors using a set of parameters and rules. For example, ScriptEase allows users to generate interaction code by adapting existing game design patterns~\cite{Gillis}. Other low-code development methods include programming by demonstration (PBD), which allows users to develop programs by demonstrating examples of desired behavior, and inductive program synthesis (IPS), which utilizes machine learning algorithms to generate programs from other example programs. For example, Rapido uses PBD to allow desktop and mobile-based AR prototyping \cite{leiva2021rapido}.

 There are also a number of tools and platforms that use NLP techniques to enable users to build applications using natural language descriptions. LLMs have demonstrated the potential to transform the way we generate, communicate, and implement ideas~\cite{dale_2021}. In particular, integration of LLMs into desktop software development processes has shown significant improvements in efficiency and accessibility~\cite{peng2023impact, poldrack2023ai}. Some GPT-based systems such as Codex~\cite{Codex} have been explored for primitive XR world-building. For example, Codex Pong uses Codex to generate code for a VR Pong game at runtime~\cite{Andrzej}. However, like many NLP-based generative programming techniques, their implementation is limited to a specific use case as it depends on the reliable generation of C\# code, which becomes increasingly complex and error-prone with the size of the script. Our work avoids this common pitfall and minimizes the risk of generating inaccurate code by abstracting functionality to a higher-level language that is optimized for natural language based generation.

Generally, AtomXR focuses on addressing two key barriers in XR development process identified by Ashtari et al.~\cite{ashtari2020creating}: 1) difficulty knowing where to start and 2) too many unknowns and changes in development, testing, and debugging. AtomXR was designed to both make XR development faster and easier for beginners to get started, and abstract away fragmented and moving parts in the complex development pipeline to reduce unknowns.

\subsection{Natural User Interfaces}

Natural language interfaces mediating human-computer interaction have been explored since the early days of interface-based computing. Notably in 1980, researchers at MIT demonstrated "Put-That-There", a natural language and gesture interface that allowed users to control the position of simple shapes on a large screen~\cite{PutThatThere}. From there, natural language interfaces have become increasingly complex and powerful, finding major use cases in voice assistant technologies ~\cite{kepuska2018next}, such as Apple’s Siri~\cite{bellegarda2013spoken}, Amazon’s Alexa, and Microsoft’s Cortona. These voice assistants perform anything from sending messages to searching the Internet to interfacing with smart home devices. However, natural language interfaces have been under-explored within XR development, largely because of the difficulty of accurately interpreting speech into actions in complex development contexts \cite{fernandez2020developing}.

On the other hand, gesture-based interfaces have advanced alongside advancements in sensing and video processing technology, with advanced hand tracking and gesture recognition becoming standard within XR headsets. We combine natural language and gesture-based interfacing techniques to achieve accessibility, naturalness, and economy of expression within our application development system.

\section{Design Goals}

\subsection{Current Development Process and Pain Points}

The following scenario illustrates how Alex, an engineer who is unfamiliar with XR development, would attempt to build a simple coin collision game using Unity, a commonly used desktop software for XR development.

Alex first sets up the Unity project, navigating through project settings, downloading the appropriate XR packages, and adding scene configuration game objects to the scene so that it is compatible with his headset, the HoloLens 2. After the setup process, Alex adds a coin to the scene by searching for, downloading, and importing a model from an online database. As he is on a 2D screen, uses the mouse and keyboard to pan, rotate, and zoom to position the coin in the scene. To add functionality, Alex adds a new script component to the coin object. To implement a collision detection script, Alex first reads Unity's documentation to understand different listeners and game object properties (rigid bodies, dynamic and kinematic types, the difference between variations of the collision listener function) before he can implement the right listener function. After he writes the script, he navigates back to the Unity interface to add and adjust components to the collision objects. 

Alex then presses play to test out his script and uses the WASD keys and the mouse to move around in the simulator. After the simulator test works, Alex wants to make sure that this works inside the headset, so he sets up Unity to live stream into the headset. After pressing play in the headset, Alex realizes that the coin is too small when deployed to the headset (and not on a 2D screen), so he returns to the desktop to adjust the size of the coin and continues this cycle of development for the rest of the game.

Based on this user journey, we identify three major pain points in the XR prototyping process:
\begin{itemize}
\item \textbf{Pain Point \#1: Complex Workflow.} The current development process requires multiple steps, waiting periods, and environments across which developers must context switch both mentally and physically (e.g. setting up a new project in Unity, installing packages and finding assets online, programming in a separate IDE, testing on the target device), which contributes to a steep learning curve and general development friction.
\item \textbf{Pain Point \#2: Programming.} Users need to program and debug complex C\# scripts, which is prohibitive for those without prior programming experience and can be inefficient for those with programming experience trying to rapidly prototype.
\item \textbf{Pain Point \#3: Development-Testing Environment Mismatch.} The current development workflow lacks support for immersive iteration where users can see and test what they are building in 3D as they are building it. Immersive world building tools like ShapesXR and FigminXR lack the ability to create game logic, separating the development process for design and functionality of applications.
\end{itemize}

\subsection{Design Objectives}

In this work, we explore how natural language combined with physical inputs can improve the XR prototyping experience by both addressing the above-mentioned pain points and opening up new possibilities for interaction. We developed the AtomXR system with the following three design goals:


\begin{itemize}
\item \textbf{D1: Natural interaction via multimodal intent recognition.} The system should allow users to directly and naturally express intention with natural language input and physical interactions to allow for ease of learning and ease of use. 

\item \textbf{D2: Easy logic design via programming abstraction and NLP.} The system should minimize intent-to-input translation by converting user-described logic into concise code (AtomScript) with transparency to facilitate error correction. 
\item \textbf{D3: Rapid immersive iteration via streamlined in-headset environment.} The system should enable users to iterate rapidly while inside the headset, allowing for immediate immersive testing without context switching. 
\end{itemize}


\section{AtomXR System}


In this section, we describe in detail the features, architecture, and implementation of the AtomXR system.

\begin{figure}
  \includegraphics[width=0.9\linewidth]{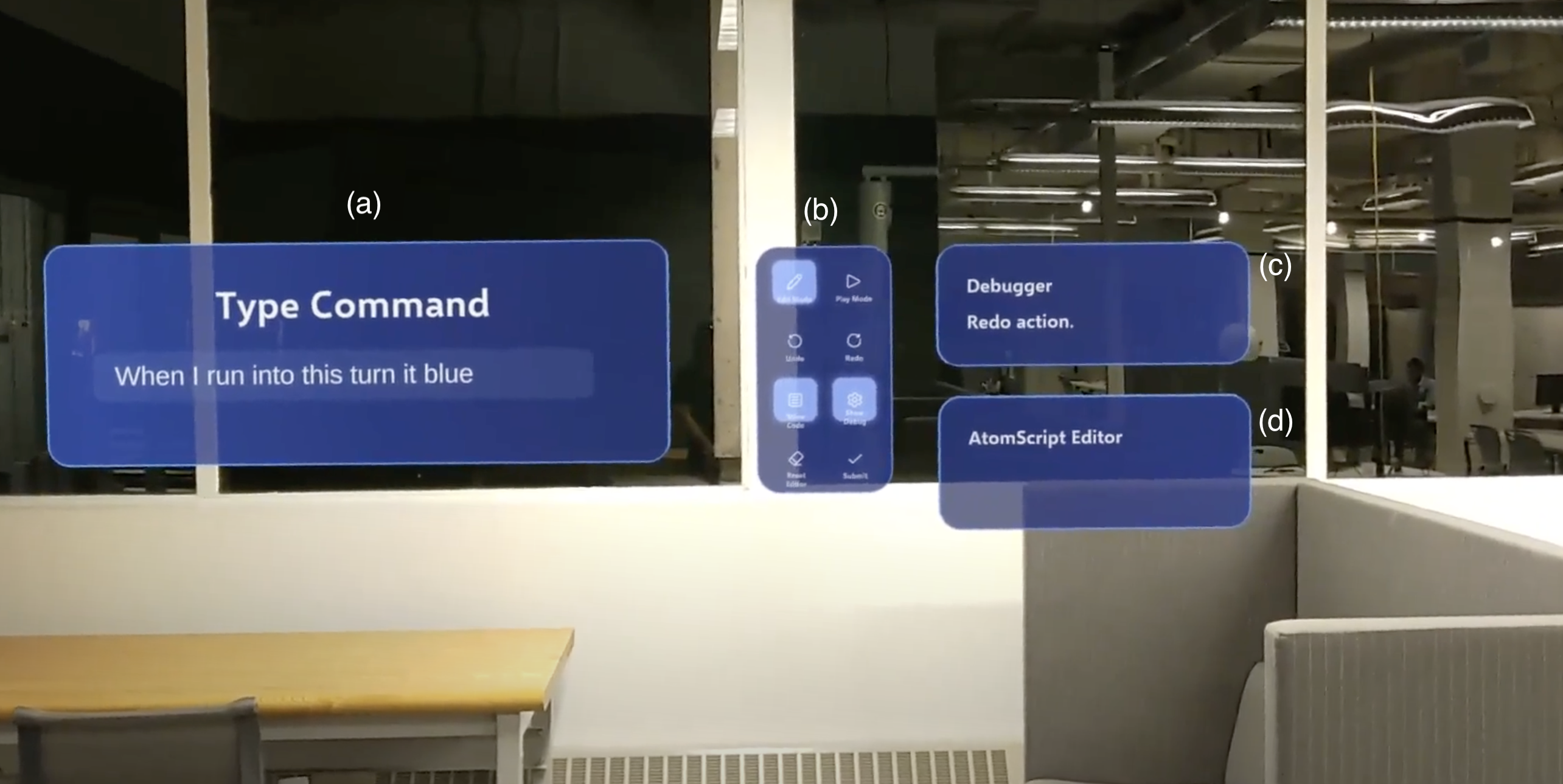}
  \caption{The four main components of AtomXR's user interface. (a) The Type Command Box allows the user to input natural language commands; (b) the menu allows toggling between edit mode and play mode and performing global functionality; (c) the debugger Panel allows logging errors; (d) AtomScript Editor allows viewing and deleting the generated AtomScript code.}
  \label{fig:test}
  \Description{A full view of the AtomXR user interface. On the left, there is a Type Command panel containing the example input, "When I run into this turn it blue". In the middle, there is a menu panel that contains four rows of two buttons each (row 1 contains play and edit mode buttons, row 2 contains undo and redo buttons, row 3 contains buttons to toggle the visibility of menu panel, and row 4 contain reset game and save game buttons). On the right, there is a debugger panel that displays system feedback to the user, and underneath that, the AtomScript Editor panel is used to display the generated code.}
\end{figure}

\subsection{System Components \& Architecture}
The top-level primitive of AtomXR is a scene. Every scene contains objects and scripts. 

\subsubsection{Objects}
Objects are 3D models in the scene. Objects are created based on natural language requests. The 3D models are sourced from a pool of built-in assets. If the requested object does not exist in the built-in asset pool, AtomXR searches an online database to find the closest related model. Each object has several attributes: size, orientation, location, and color. These attributes are represented in an object information panel that hovers above the object in the editor view.
Each object also has methods for updating object attributes or deleting the object. These methods are triggered by natural language commands (e.g. \textit{"Make this larger", "Turn this blue"}) and physical interactions (grabbing an object and moving it, pressing the delete button, etc.).

\subsubsection{Scripts}
Scripts are pieces of AtomScript code that define the behavior of the scene and its objects. AtomScript code is generated using our NLP system, which translates natural language to AtomScript. Scripts are interpreted at runtime by the AtomScript interpreter built in Unity. Methods include but are not limited to creation and updating of variables, for and while loops, on start logic, if then logic, mathematical operations, playing sounds, runtime object attribute changes, runtime object creation and deletion, and specialized 3D application functions, such as collision detection and button press detection.

\subsubsection{Global Development Environment}
Beyond the elements within a scene, AtomXR also serves as a global development environment (Fig.~\ref{fig:test}), including:
\begin{itemize}
\item Play and edit modes, which allow users to alternate between adding functionality in an information-rich display of the scene and testing how the final application will behave.
\item Ability to view and delete generated scripts. 
\item Undo and redo functions, which allow users to revert to a previous or future state.
\item Reset function, which allows users to erase the current scene and restart the development process.
\item Save function, which allows users to save their application specification to a cloud database.
\item Hybrid eye gaze, gesture, and natural language interaction methods to create, manipulate, and reference objects.
\end{itemize}

\subsubsection{System architecture}
\begin{figure}
\centering
\includegraphics[width=.85\linewidth]{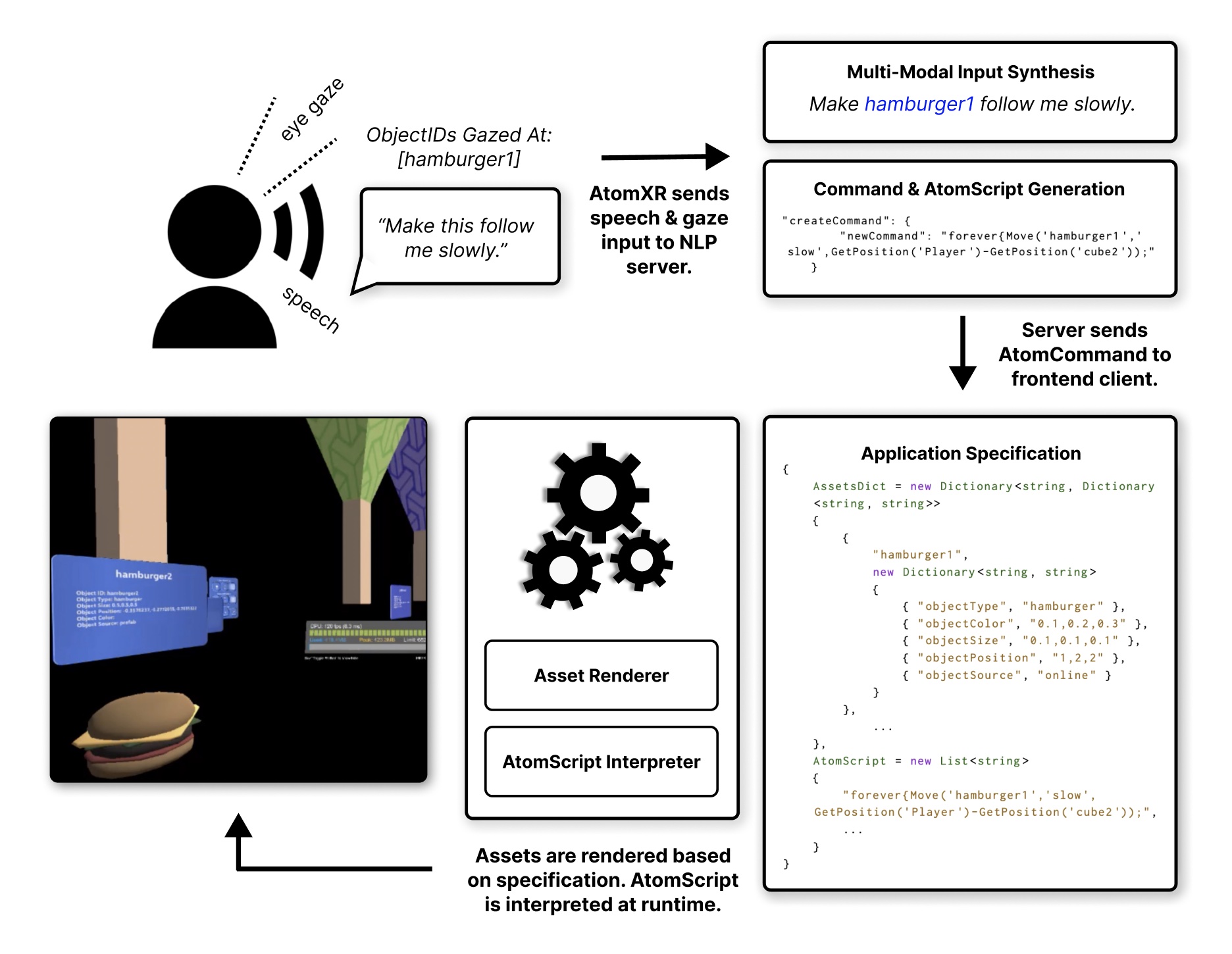}
\caption{After a user gives AtomXR a verbal command, we use the HoloLens built-in speech-to-text engine to convert this to text. This text, along with the user's eye gaze targets while speaking, is sent to an online webserver. The webserver uses the GPT-3 API and semantic embeddings to synthesize the inputs and convert this text into a command that defines how to edit the application specification, including adding new AtomScript code. This application specification is then rendered and AtomScript code is exected on the user's headset at runtime.}
\label{fig:process}
\Description{Flow chart demonstrating the system processes for AtomXR. In the top left, an icon of a user speaking is shown, with the conversational text box containing, "Make the tree throw hamburgers at me when I'm running around". An arrow connects this section to the top right section, with the caption under the arrow, "AtomXR sends web request with speech-to-text dialog to the server." The top right section contains a screenshot of the AtomScript generated by GPT3, labeled, "Convert verbal command to AtomScript command". An arrow connects this section to the bottom right section, with the caption next to the arrow, "Server responds to web request with AtomScript command." The bottom section contains a gear icon labeled, "The AtomScript code is executed by the AtomScript interpreter". An arrow connects this section to the bottom left section, which contains a screenshot of the game with the trees and hamburger.}
\end{figure}

The AtomXR system, diagramed in Fig.~\ref{fig:process}, consists of a frontend development environment and a backend web server. The immersive frontend application supports D3 and was built in Unity and deployed to Microsoft's HoloLens 2. We use the built-in text-to-speech functionality of the HoloLens keyboard to transcribe what the user is saying, and use the eye-tracking and hand-tracking data from Microsoft's Mixed Reality Toolkit (MRTK) to enable our eye-gaze and touch interactions. This multimodal input data is then sent to a persistent backend web server runs on the cloud via Amazon Web Services (AWS). The server NLP system, described in detail in Sec.~\ref{sec:nlp}, synthesizes the several input streams and returns a user intention interpretation that is then used to update the specification of the application. The specification of an application follows a nested json format, an example of which is shown for a simple chase game in Fig.~\ref{fig:process}. This specification is used to render assets and execute AtomScript code at runtime.

Below, we described the implementation details for key components of AtomXR, including the user interface (Sec.~\ref{sec:interface}), interaction scheme (Sec.~\ref{sec:interaction}), AtomScript (Sec.~\ref{sec:atomscript}), and NLP system (Sec.~\ref{sec:nlp}).


\subsection{User Interface}
\label{sec:interface}

\subsubsection{Type Command Box}
The Type Command box (Fig.~\ref{fig:test}a) allows users to input natural language requests to build their application. Pressing on the input box will open up a virtual keyboard. Here, the user can either use their hands to manually type into this virtual keyboard, or they can press the microphone button on the keyboard to speak their request. We include both input methods but recommend the microphone input as typing using a virtual keyboard in augmented reality is less efficient \cite{derby2019text}.

\subsubsection{Menu}
The first row of the menu panel (Fig.~\ref{fig:test}b) contains a toggle between AtomXR's two modes: edit and play mode. In edit mode, users can use natural language to create their application, view the debugger (Fig.~\ref{fig:test}c) and the current AtomScript code (Fig.~\ref{fig:test}d), and manipulate the properties of objects. In play mode, users can test the application logic from the end user's perspective. Users switch between edit and play modes often when developing their applications. In the second row of the menu, users can undo and redo actions that have been taken in edit mode. In the third row, users can toggle the visibility of the Debugger panel (Fig.~\ref{fig:test}c) and AtomScript Editor (Fig.~\ref{fig:test}d). In the final row, users can press the reset button to restart their development process, or the submit button to save their game to a persistent cloud database.

\subsubsection{Debugger Panel} (Fig.~\ref{fig:test}c) The debugger panel is used as a feedback mechanism to update users on the system status (e.g. when the server is processing requests, when errors arise, etc.).

\subsubsection{AtomScript Editor}
The AtomScript Editor (Fig.~\ref{fig:test}d) shows all of the currently written AtomScript code. Each segment of code will appear as a separate block. Users can delete code by pressing the delete button on the code block. Due to the difficulties of typing in AR, we do not allow users to edit code but instead encourage users to just delete the code and re-try their voice command. For future work, we expect to trial more manipulable code representations rather than just plain text, such as a block-based system similar to MIT's Scratch platform.

\subsection{AtomXR Interaction Design}
\label{sec:interaction}

To support D1 and enable D3, we developed an interaction scheme integrates several natural modalities of user input (natural language, eye gaze, and physical interactions) to allow users to directly express intention without being restricted to a single, pre-defined method of communicating with the system. 




\subsubsection{Natural Language Interaction}
The primary method for a user to make requests to the system is through natural language, inputted either through dictation or typing. Voice commands are often faster and more natural for conveying intent than the standard keyboard. However, speech-to-text accuracy issues sometimes makes it slower, so the standard keyboard is also available to the user. Users can speak to the system conversationally as if they were speaking to an assistant. Users can create (e.g.,\textit{"Create a small red cube"}) , update (e.g., \textit{"Make the cube blue"}), and delete objects (e.g., \textit{"Delete this cube"}) and specify logic using natural language (e.g., \textit{"If I run into this cube, turn it green" or "Make this cube follow me slowly"}).
To create a simple chase game, the game designer could say \textit{"Give me a turtle", "Make the turtle chase me", "Make the turtle move faster if I start moving faster".}

This natural language interaction accommodates the many ways a user may choose to express the same intention. By using LLMs and semantic embeddings to convert natural language into code, variations in language with the same intent yield the same code. Users can express ideas without matching a specific syntax. For example, to create an object into the scene, \textit{"create a cube"}, \textit{"generate a 3-dimensional square"}, \textit{"put a cube into the scene"}, and \textit{"can you make me a box"} all result in the same cube being placed in the scene. This flexibility to linguistic variation makes our natural language interaction method syntax agnostic, where users can directly express their thoughts instead of learning to map their thoughts into a specific accepted programming syntax. The direct expression of thought allows users to focus on creating the substance of an application rather than transcribing substance into specification. The NLP backend that supports these interactions is described in detail in ~\ref{sec:nlp}.

\subsubsection{Eye Gaze Interaction}
Eye gaze interaction, enabled by the eye-tracking capabilities of the HoloLens, provides supplemental information to support natural language interaction, allowing the user to leave ambiguities in their language by predicting intent from eye-tracking. Specifically, eye gaze helps users reference objects when giving natural language commands to specify which objects to apply the command to. Instead of referring to objects by their object ID, a user can use a demonstrative pronoun while gazing at the object- for example, instead of saying \textit{"Make \textbf{cube1} orange"}, a user can say, \textit{"Make \textbf{this} orange"} while gazing at the cube. This works for any kind of reference substitution, including updating objects
(e.g., \textit{"Make \textbf{this} larger"}), deleting objects
(e.g., \textit{"Delete \textbf{that}"}), and defining logic involving one or multiple objects 
(e.g., \textit{"Make \textbf{this} (gaze at object 1) revolve around \textbf{that} (gaze at object 2)"}).
The system provides validation for the user by darkening the color of objects when the user gazes at them, indicating that it has detected their reference to that object. This interaction workflow mimics what people naturally do when communicating with others, again reducing the effort required from the user to match their natural expressions of intention with system input methods.

\subsubsection{Physical Manipulation}
We implemented hand-gestural interactions based on the hand-tracking capabilities of the HoloLens to complement the natural language and eye-gaze interactions for physical manipulation.
%
Users can easily resize, reposition, and delete virtual objects through intuitive hand gestures like pinching and dragging. This in-headset object manipulation enables a real-time, 3D perspective view of the real application, which offers a significant advantage over traditional 2D interfaces that lack depth and scale awareness. Users can also interact with virtual buttons and control menu panels with simple hand movements.

\subsection{AtomScript}

To support D2, we designed AtomScript, a high-level programming language that AtomXR generates to define game logic. AtomScript had two design goals. First, the language should implement the core functionality necessary for developing XR applications, including both fundamental programming constructs and functionality specific to 3D and XR applications. Second, the language's syntax should be as semantically close to human thought as possible, such that it is both easier for the human user to understand and conducive to generation from natural language via LLM. 

AtomScript abstracts away the low-level complexity of the underlying Unity C\# on which it was built. Directly generating C\# code from natural language has been found to be prone to error and difficult to scale when asked to generate increasingly copmlex code \cite{Andrzej}. This is due to the large number of syntax and usage rules that must be followed to generate working code. AtomScript improves the quality of converting natural language to code by following a simpler syntax. While the current version of AtomScript is much less capable than Unity C\#, for rapid prototyping of XR experiences, AtomScript consolidates the largely unused optionality that other game engines like Unity and Unreal provide. Through this design, LLMs can effectively generate AtomScript code more accurately and humans can understand it better. The grammar of AtomScript is detailed in \ref{sec:atomscriptgrammar}

Below, we introduce AtomScript's core features and properties.

\subsubsection{AtomScript Core Features}
\label{sec:atomscript}

\noindent \textbf{Basic elements.}
AtomScript has three fundamental types: strings, numbers, and arrays. Variables can be declared and assigned to each of these data types. These data can be passed into AtomScript's built-in functions.

\noindent \textbf{Built-in functions.}
To align with the capabilities of most 3D game engines, AtomScript supports essential operations like object creation, deletion, and manipulation. It also features an event-handling system and separate functions for code that initializes at game start and code that runs continuously each frame.

Based on established game design patterns, AtomScript provides several built-in listeners that allow the user to perform certain operations on specific game events, including 
\textit{onStart}, which runs at the start of the game; \textit{forever}, which runs every frame; \textit{onButtonPress}, which triggers when a user presses a "button" object; \textit{onCollision}, which triggers when two game objects collide. 
To allow rapid prototyping for common XR applications, AtomScript also features a number of built-in functions. These functions are the "atomic" building blocks of more complex game interactions. These include functions like \textit{Move}, \textit{ChangeColor}, \textit{GetPosition}, and \textit{PlaySound}, among several other built-in functions.

We demonstrated how users use AtomScript to prototype realistic applications in Sec.~\ref{sec:UsageScenarios}.

\subsubsection{AtomScript Properties}
The design of AtomScript provides the following properties as a prototyping language in comparison to traditional Unity C\#. \\
\noindent \textbf{Completely text-based}: 
To enhance the interpretability of prototyping scripts, the generation of AtomScript is purely text-based. This allows AtomScript to be stored completely as human-comprehensible text and could be easily generated by AI. Many interactions in game engines like Unity and Unreal Engine are carried out using a UI rather than code, but using text as the authoring interface allows the entire game space to be within the output space of state-of-the-art generative text AI. 

\begin{figure}[h]
\centering
\includegraphics[width=\linewidth]{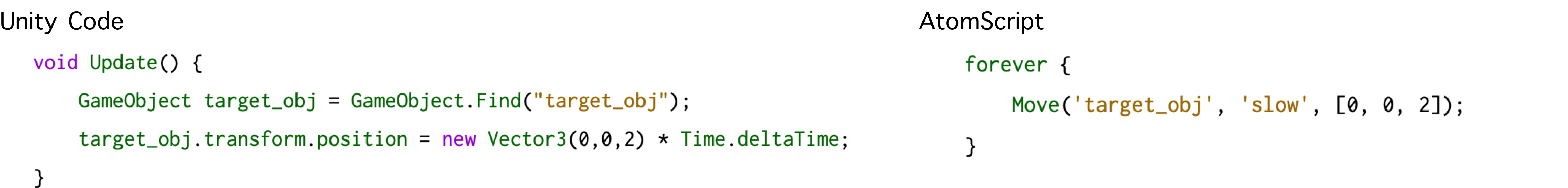}
\captionof{figure}{Scripts for moving an object in Unity and AtomScript.}
\Description{An image showing two scripts for moving an object in Unity and AtomScript.}
\label{fig:move-obj}
\end{figure}

\noindent \textbf{Simple, high-level interface}: 
To enhance prototyping for XR applications, AtomScript abstracts away complex functions to improve usability and learnability by non-programmers. 
AtomXR prioritizes user experiences over robustness with a focus on rapid prototyping.
%
For example, we optimized routine operations in XR experiences like object movement for conceptual simplicity to enhance user experience.
As shown in Fig.~\ref{fig:move-obj}, to move an object in a straight line, a developer would write the following in Unity C\# (left), to replaces the target\_obj from its previous position by 2 in the z-axis. 

While such low-level abstraction may be suitable for many complex, non-XR projects, it’s unnecessary for most XR applications and can even be difficult for novice developers.
The simplified expression of AtomScript (right) allows the user to evaluate the correctness of the AI-generated code blocks even if they lack the expertise to write the code themselves.



To further improve understandability, we made certain design decisions like naming the position command as "GetPosition" instead of "transform.position" and the vector declaration as [0, 0, 2] instead of Vector3(0, 0, 2).


\noindent \textbf{Limitations.}
Although AtomScript is engineered for simplicity and ease of use, it does have its limitations compared to Unity C\#. For instance, it lacks the ability for users to create custom functions and doesn't offer fine-grained control over interactions such as collision detection and object attributes.

These features are often crucial in applications requiring detailed control. However, it's worth noting that while Unity is tailored for multi-platform, production-level gaming, AtomScript is purposefully designed with a narrower focus: to facilitate rapid XR prototyping and development.





\subsection{Natural Language Processing}
\label{sec:nlp}
To support D1 and D2, we developed an NLP system that combines LLMs with semantic embeddings to interpret user intent from multimodal input.

\subsubsection{Intention Recognition}

Each time a user submits a request, their natural language input is sent to the NLP server along with information about their eye gaze (see first process arrow in Fig.~\ref{fig:process}). Few-shot prompted GPT-3 is then used to generate an AtomCommand, which is a JSON object specifying a modification of the game state.\footnote{AtomXR was built in 07/2022, predating GPT-4}. Example syntax of AtomCommands for creating logic, and creating, updating, deleting objects are shown below.

\begin{figure}[h]
\centering
\includegraphics[width=.8\linewidth]{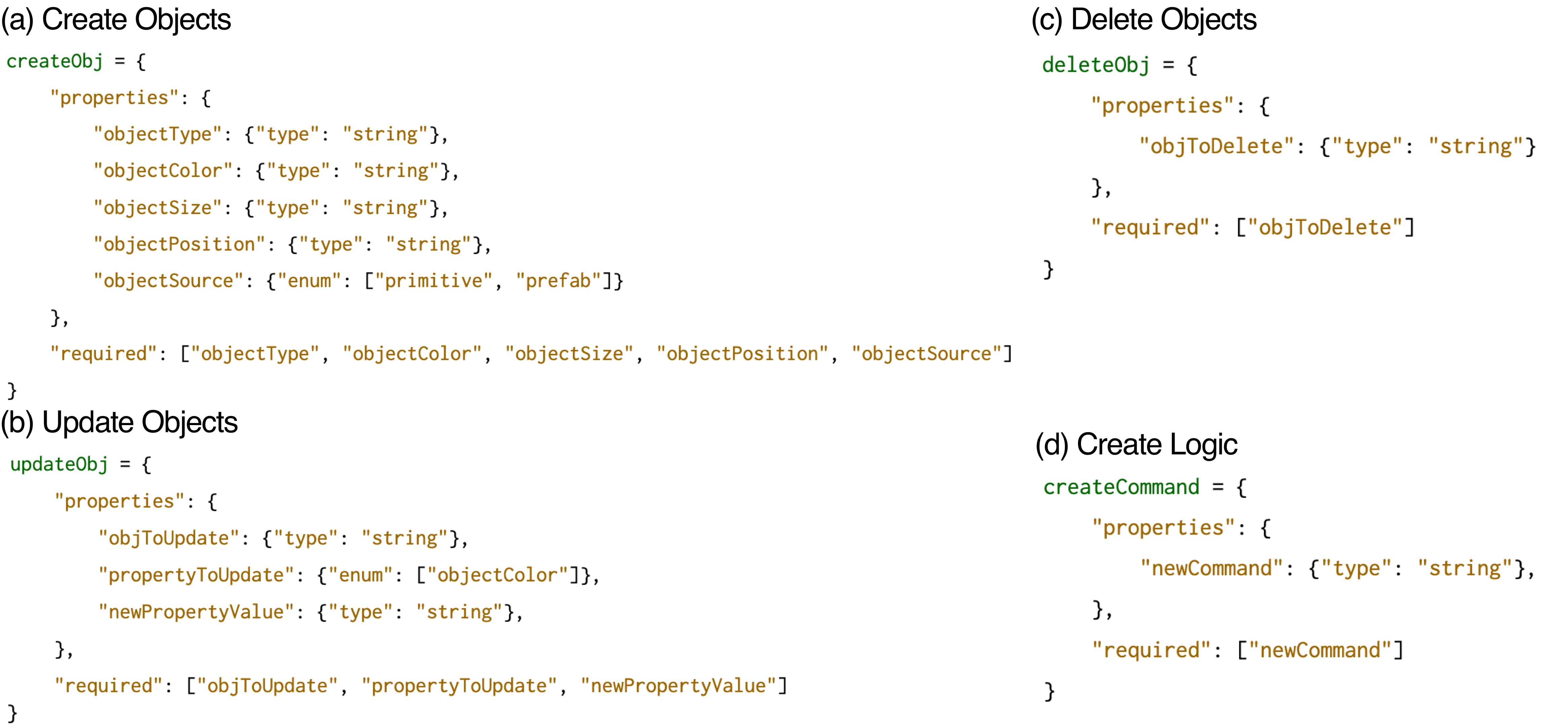}
\captionof{figure}{The expected format of an AtomCommand. AtomCommands are generated by GPT-3 and specify the AtomXR game state. }
\Description{The expected format of an AtomCommand. AtomCommands are generated by GPT-3 and specify the AtomXR game state. An image showing 4 examples of AtomCommands.}
\label{fig:command}
\end{figure}

These AtomCommands are sent to the Unity frontend application and executed (see second process arrow in Fig.~\ref{fig:process}). Using LLMs to convert natural language (e.g. "create a red ball") into an AtomCommand is generally accurate when classifying the type of command (i.e. logic creation and object deletion/creation/updates). The key challenge in the NLP process is generating the AtomScript code that matches the intent of the author. 

\subsubsection{Generating AtomScript}

\lstdefinestyle{plainblackstyle}{
    basicstyle=\ttfamily\small,   
    breakatwhitespace=false,      
    breaklines=true,             
    numbers=none,                 
    showspaces=false,           
    showstringspaces=false,
    showtabs=false,             
    tabsize=2,
    backgroundcolor=\color{white},  
    commentstyle=\color{codegreen},     
    stringstyle=\color{codegreen},      
    identifierstyle=\color{codegreen},
    ndkeywordstyle=\color{codegreen},
    keywordstyle=\color{codegreen}     
}

\lstdefinestyle{g4style}{
    basicstyle=\footnotesize\ttfamily,
    keywordstyle=\color{blue},
    commentstyle=\color{gray},
    stringstyle=\color{codepurple},
    keywordstyle=\color{codeorange},
    numberstyle=\color{codegray},
    identifierstyle=\color{codeblack},
    showstringspaces=false,
    breaklines=true,
    tabsize=2,
    captionpos=b,
}

To generate AtomScript, we prompted GPT-3 with example pairs of natural language and AtomScript in the following format:
\begin{lstlisting}[style=g4style]
SPEECH:Make the octopus5 disappear and play a noise after 4 seconds
ATOMCOMMAND:{"createCommand":{"newCommand":"currTime=TimeSinceStart();
forever{if(TimeSinceStart()-currTime >= 4) {Disappear('octopus5');Play('noise');}}"}}
###
SPEECH:When the variable building is equal to 3, then make the table3 play a noise
ATOMCOMMAND:{"createCommand":{"newCommand":"forever{if(building==3){PlaySound('table3');}}"}}
###
...
\end{lstlisting}

We took a trial and error approach to prompt engineering, correcting for edge case errors as they arose. 

\subsubsection{Approximation Using Embeddings}

When a user requests an object, they may use a number of different names to refer to the same kind of object. For example, a user may ask for a "red apple", a "fresh apple", or simply an "apple". AtomXR contains a pre-loaded database of 3D models, including built-in Unity primitives (cube, sphere, capsule, etc.) and additional 3D models we curated (fruits, collectible items, household items, parts of nature, etc.). If the name of the object requested by the user does not directly match the name of an object in the AtomXR database, we use a semantic embedding calculation (based on TensorFlow's universal sentence encoder \cite{cer2018universal}) to determine whether there exists a similar object in the database or whether we should query the object from an online database. Specifically, if the embedding vector of the object name requested by the user is close enough to that of an object name within our database (> 0.75 cosine similarity), we use the model from our database. If not, we query the online Sketchfab database to find relevant models. This method provides approximations of user intent that allow the user flexibility in describing desired objects.

\subsubsection{Referencing Objects}
Our eye gaze referencing system that allows users to use ambiguous language when referring to objects (\textit{"Make \textbf{this} blue"}, \textit{"Delete \textbf{that}"}, etc.) is enabled by a few-shot prompted GPT-3 call that replaces ambiguous demonstrative pronouns with the objectID of the object the user gazes at. While this processing could also be implemented through simple part-of-speech (POS) tagging, we used an LLM method to avoid edge case errors. The few-shot prompt is formatted as follows:\\

\begin{lstlisting}[style=plainblackstyle]
Replace the appropriate words in the string:
STRING: If I touch this blue box then change this box's color.
OBJECT LIST: box12, box23
NEW STRING: If I touch box12 then change the color of box23.
###
Replace the appropriate words in the string:
STRING: If this sphere runs into that cylinder then increase the scoreboard variable by one.
OBJECT LIST: sphere10, cylinder38
NEW STRING: If sphere10 runs into cylinder38 then increase the scoreboard variable by one.
###
...
\end{lstlisting}

\section{Usage Scenarios}
\label{sec:UsageScenarios}





To demonstrate the expressiveness of AtomXR concretely, we present two example experiences authored by Alex using the AtomXR system. Alex serves as a representative adult with limited coding skills and minimal familiarity with AR environments, yet has a keen interest in prototyping XR experiences. 

\subsection{Space Shooter}

The following scenario illustrates how Alex can use AtomXR to quickly build an augmented reality experience using natural language and physical interactions inside his headset. Alex wants to build a complex space shooter game where the player is inside a spaceship, with asteroids flying towards him. The player should be able to press 2 buttons to move the turret left or right, and another button to fire at the asteroids. The game starts with 5 asteroids hurtling toward the player, and every asteroid that is destroyed spawns 2 more asteroids, making the game increasingly hard. If the player surpasses 20 score, they win the game. 

\begin{figure}[h]
    \centering
  \includegraphics[width=\linewidth]{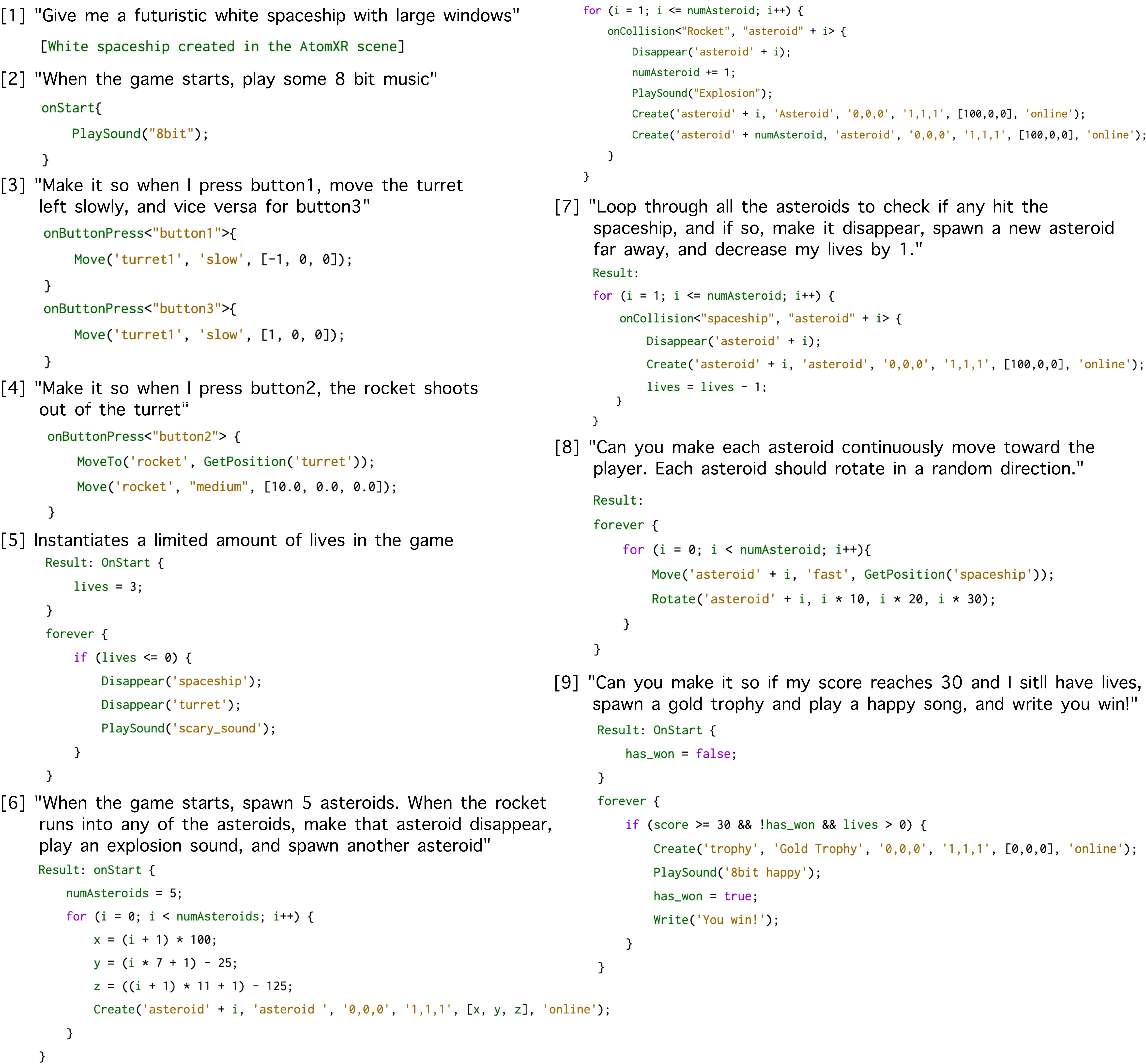}
  \caption{The prompts spoken by Alex and the AtomScript generated while creating the space shooter game.}
  \label{lst:label1}
  \Description{The prompts spoken by Alex and the AtomScript generated while creating the space shooter game. There are 9 total prompts shown, and below each prompt is the associated AtomScript.}
\end{figure}

To complete the game, Alex prompted AtomXR step-by-step, as shown in Script~\ref{lst:label1}.
We described each step sequentially while referring to the prompts and corresponding code in Script~\ref{lst:label1}.


\begin{figure}
    \centering
  \includegraphics[width=\linewidth]{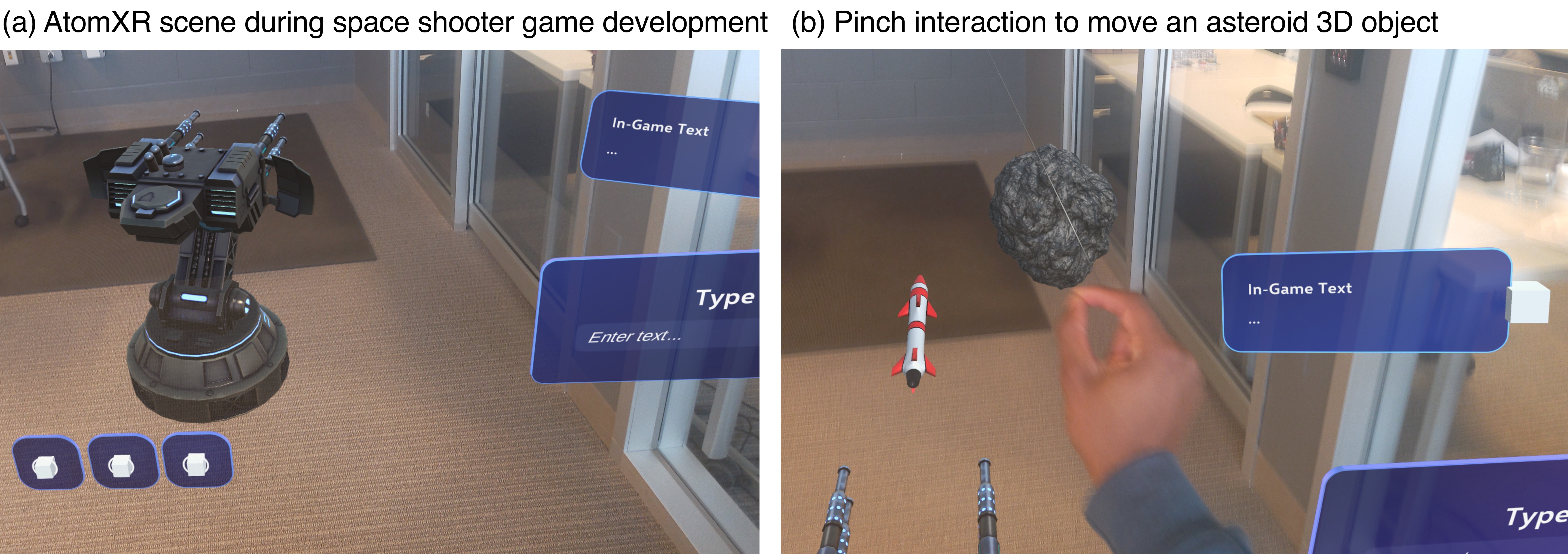}
  \caption{A space shooter game developed using AtomXR. The in-headset photos show (a) the game scene during development containing AtomXR Type Command Box, the turret, and 3 control buttons, and (b) Alex using simple and intuitive pinch gestures to  an asteroid 3D object.}
  \label{fig:space-shooter-game}
  \Description{In-headset photo of the AtomXR system in action in the space shooter example game, with the user adjusting a 3D asteroid object using a pinch gesture. Beneath the asteroid are a rocket and the barrels of the turret from the space shooter game. The game is situated in AR, where the background environment is still visible.}
\end{figure}

Alex starts prototyping with AtomXR by putting on his headset, opening the AtomXR application, and dictating into the command box,
\textit{"Give me a futuristic white spaceship with large windows"} (Fig. \ref{lst:label1}[1]). On the backend, AtomXR uses the SketchFab API to find a spaceship model matching the query. From there, Alex moves and resizes the spaceship by grabbing and dragging it around with his hands, so that the spaceship is large enough to fit him inside. Alex then asks for a \textit{"Sci-fi turret"}, a \textit{"rocket"} for firing, and 3 buttons for moving and firing the turret. These items appear into his AR world as shown in Fig. \ref{fig:space-shooter-game}a.



Alex asks for music to be added to the game (Fig. \ref{lst:label1}[2]). Alex can see the code that was generated in the AtomScript viewer/editor. 

Now, with all the components in the world and music, he starts adding turret logic to the game, which allows the user to move the rockets left and right, as well as shoot with the buttons (Fig. \ref{lst:label1} [3, 4]). 
%
Alex then goes into Play Mode to verify that the AtomXR-generated game works. When he enters play mode, the music plays. When he presses the left and right buttons, the turret moves accordingly. When he presses the middle button, the rocket fires from the turret, as expected. 

Alex decides to add a new mechanic, where he wants the user to only have a limited amount of lives in the game. He creates a system where if you lose 3 lives, then the game is over (Fig. \ref{lst:label1} [5]).
He wants the number of asteroids at the start to be 5, and when the rocket hits an asteroid, increment the number of asteroids by 1, destroy the hit asteroid, and spawn 2 more in random locations (Fig.~\ref{lst:label1} [6, 7]).
After testing this out in play mode, Alex notices the game is too easy because the asteroids are just standing in place. He asks for the asteroids to chase him (Fig.~\ref{lst:label1} [8]).
%
%
Alex again goes into Play Mode to test the functionality. Fig. \ref{fig:space-shooter-game} shows the objects in the current scene. 
Wanting the game to have a satisfying conclusion, Alex adds an end-state where the user can win (Fig.~\ref{lst:label1} [9]).
%
Entering Play Mode, Alex plays the completed game, and sees that everything is working together.

\subsection{Chase Game}
The following scenario illustrates how an interactive chase game can be built. This was the game we asked participants in user studies to build in order to compare the effectiveness of AtomXR to other systems. In this game, the player works to collect a cherry, which respawns in a new location every time the user runs into it. The player is being chased by a watermelon. The game also integrates several audio interactions: background music is played continuously, when the player collects the cherry, a coin collection sound is played, and when the watermelon runs into the player, a scary sound is played, indicating failure. 

\begin{figure}[h]
    \centering
  \includegraphics[width=.8\linewidth]{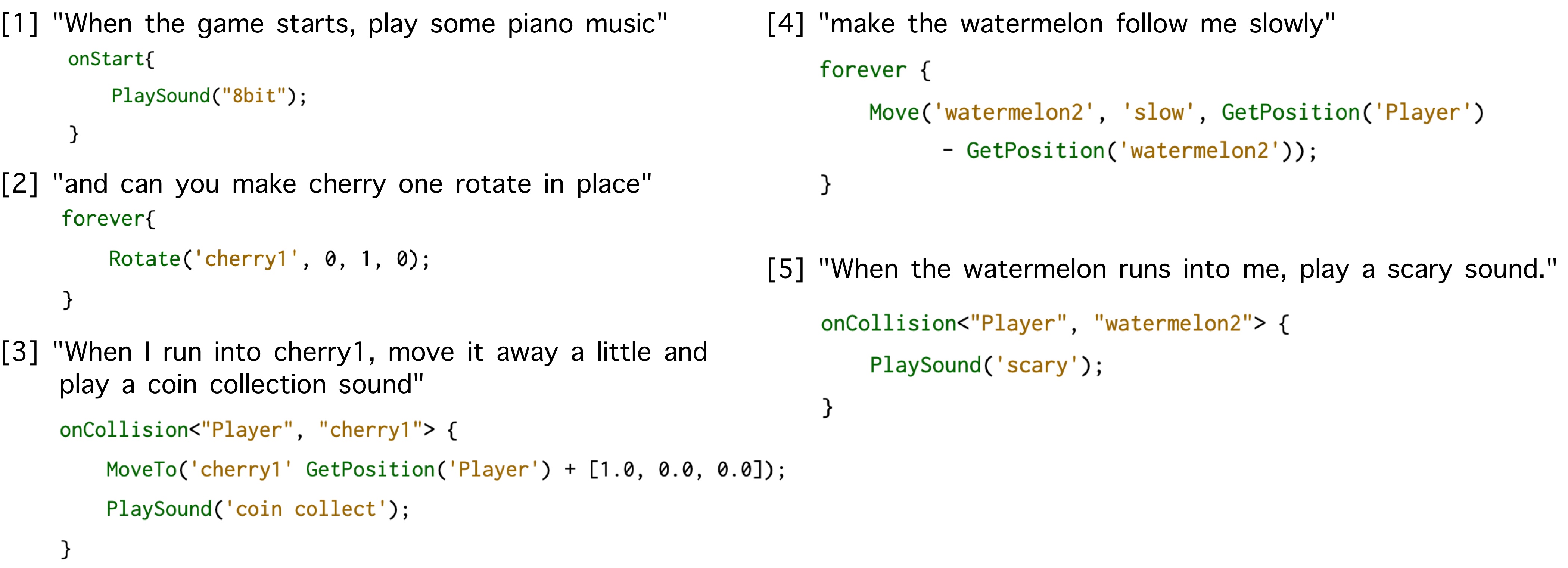}
  \caption{The prompts spoken by Alex and the AtomScript generated while creating the chase game.}
  \label{lst:label2}
  \Description{}
\end{figure}
The user starts by stating adding music to the game (Fig.~\ref{lst:label2} [1]).
\begin{figure}[t]
\centering
\includegraphics[width=1\linewidth]{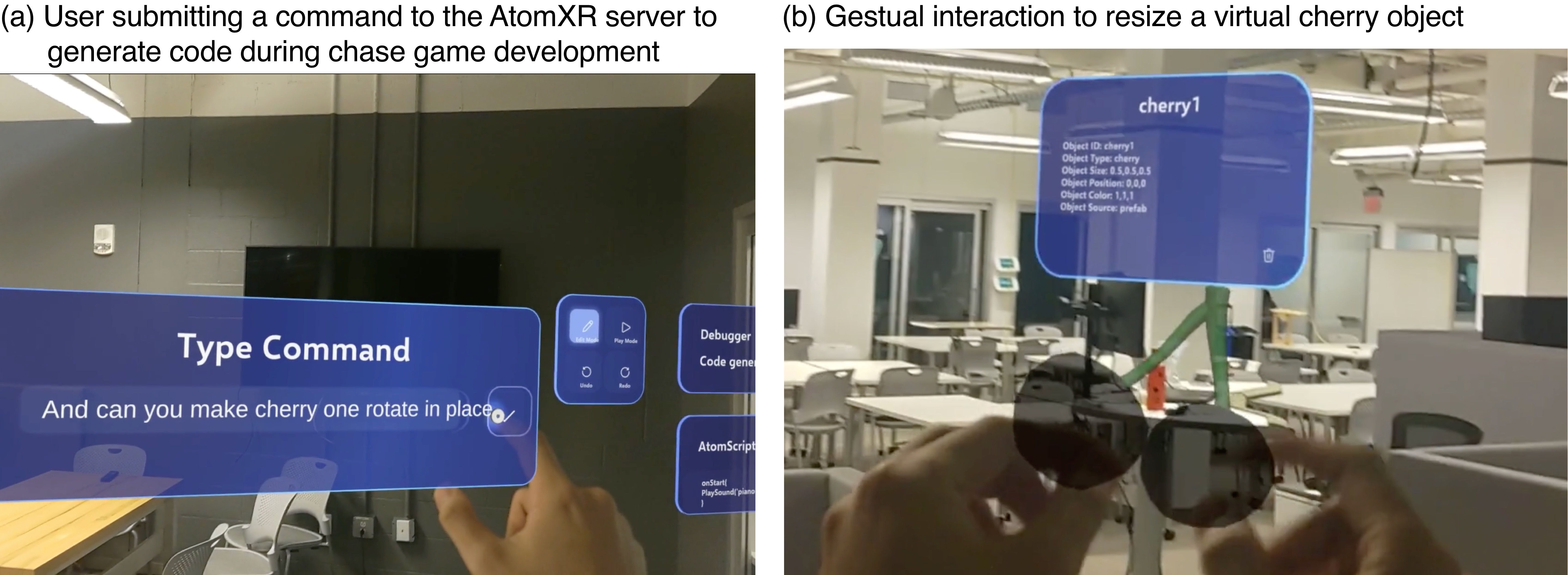}
\caption{(a) The user submitted a command to the AtomXR server to generate code while making the Chase Game. (b) The user used hand interactions to resize a cherry virtual object. Above the cherry is an info-card about its properties, including ID number, object type, object size, position, and the source of where the 3D object was spawned from.}
\label{fig:chase-game}
\Description{In-headset photo of the AtomXR system during the chase game example. The user is using a pinch gesture with both hands to resize a virtual cherry object. Above the cherry object, an information panel displays its ID and properties. The game is situated in AR, where the background environment is still visible.}
\end{figure}
%
%
The user then states: "Give me a small cherry", and the cherry spawns in their scene. It is still a bit too big and isn't in the right spot, so Alex uses his hands to grab, resize, and move the cherry to the right place, as seen in Fig. \ref{fig:chase-game}b. He then says "and can you make this cherry rotate in place" (Fig. \ref{fig:chase-game}a). Because Alex was looking at the cherry while saying "this cherry", AtomXR's eye-gaze system understands that he's referring to the recently created cherry.
Alex enters play mode and finds that the cherry is behaving as expected. He then moves onto more complex functionality, first asking for the cherry to move away and play a sound every time he collects it (Fig.~\ref{lst:label2} [3]). 
With the cherry collection functionality completed, the user creates a watermelon that chases the user around as they collect cherries (Fig.~\ref{lst:label2} [4]).

Finally, the user encodes the end-game where the watermelon plays a scary sound effect on catching the player. ( [5])
Testing the game, Alex sees the watermelon running after him slowly, and the cherry runs away every time it is collected, with coin collection sounds, scary sounds, and background music playing.






\section{Study \#1: AtomXR vs. Current Workflow}

To understand how well our system supports XR prototyping tasks in comparison to current development systems, we conducted a study in which participants were asked to complete a series of development tasks using AtomXR (in both desktop and headset variations) and the traditional Unity-based development system. 
Specifically, the study aimed to explore the following research questions:
\begin{itemize}
    \item \textbf{RQ1:} \emph{How does AtomXR compare to traditional development methods in terms of learnability, ease of use, and efficiency for prototyping interactive XR applications?}
    \item \textbf{RQ2:} \emph{How does AtomXR affect the user's subjective experience of difficulty and satisfaction during the development process?}
\end{itemize}

We performed analyses on 1) completion metrics (evaluated by experimenters, including the number of tasks completed, time taken per task), and 2) user experience metrics (evaluated by participants, including ease-of-use, learning curve, enjoyment, physical comfort).

\begin{table*}
  \caption{Experiment Tasks} 
  \label{tab:task}
  \small
  \begin{tabular}{ccl}
    \toprule
    Task \# & Task Description & Task Category\\
    \midrule
    1 & When the game begins, play some piano music. & Audio \& Start Logic\\
    2 & Create a cherry. & Object Creation\\
    3 & Change the size of the cherry to make it realistic. & Object Property Update\\
    4 & Make the cherry rotate in place. & Object Behavior\\
    5 & Play a collection sound effect when the player hits the cherry. & Audio \& Collision Logic\\
    6 & Move the cherry forward a little when the player hits it. & Movement \& Collision Logic\\
    7 & Create a watermelon. & Object Creation\\
    8 & Make the watermelon chase the player slowly. & Object Behavior\\
    9 & Play a scary sound when the watermelon runs into the player. & Audio \& Collision Logic\\
  \bottomrule
\end{tabular}
\Description{Three column table of tasks containing the Task #, Task Description, and Task Category for 9 tasks. 1) When the game begins, play some piano music (Task Category: Audio & Start Logic), 2) Create a cherry (Task Category: Object Creation) 3) Change the size of the cherry to make it realistic (Task Category: Object Property Update), 4) Make the cherry rotate in place (Task Category: Object Behavior), 5) Play a collection sound effect when the player hits the cherry (Task Category: Audio & Collision Logic), 6) Move the cherry forward a little when the player hits it (Task Category: Movement & Collision Logic), 7) Create a watermelon (Task Category: Object Creation), 8) Make the watermelon chase the player slowly (Task Category: Object Behavior), 9) Play a scary sound when the watermelon runs into the player (Task Category: Audio & Collision Logic).}
\end{table*}

\subsection{Study Design}

We designed a hybrid within-subjects and across-subjects study with three conditions, in which participants were given 20 minutes to complete a series of development tasks using 1) the Unity development system, 2) AtomXR running on desktop, and 3) AtomXR running inside the headset. Each participant experienced the Unity condition and one of the AtomXR conditions, with condition assignment and order counterbalanced to control for learning effects. We included the desktop version of AtomXR in order to disentangle the effects of interfacing with the headset from the effects of the use of natural language on the development experience. Data was collected via two post-condition and one post-experiment survey and interview.

\subsection{Participants}

We recruited 8 participants (6 female and 2 male) from university and public interest group channels, selecting to diversify experiences and attitudes toward technology.
On average, participants held positive attitudes toward new technologies (mean = 4.25, SD = 0.71, where 1 is very negative and 5 is very positive). On a scale from 1 (layman) to 5 (expert), participants on average reported a 3.13 level of prior programming experience (SD = 1.13), a 2.13 level of prior game development experience (SD = 0.83), and a 1.75 level of prior experience working with AR/VR (SD = 0.71).

\subsection{Procedure}

Participants were first given an introduction and overview of the experiment, followed by a consent form. Participants then received a 10-minute guided walkthrough of how to use the first development system in their assigned treatment group. This walkthrough covered the basics of the development system and demonstrated how to create objects, update their size and position, track collisions, and play audio. Participants were allowed to ask questions freely. Following the tutorial, participants were given 20 minutes to complete the experiment task using the first system.

We designed an integrative task in which participants were asked to complete a series of smaller tasks to build an AR game involving collecting items while evading an enemy. The task was designed to cover a breadth of different types of development tasks, from object creation to logic implementation. Table~\ref{tab:task} details the task descriptions and categories. 

We recorded the time participants needed to complete each task. After the time expired, they answered survey questions about their task experience. They then repeated this process for the second system in their assigned group, including another walkthrough and post-task survey. 
Finally, participants completed a post-experiment survey with multiple choice and open-ended questions about their overall experience during the full experiment, evaluations of each system, and preferences.

\begin{figure}[t]
\centering
\includegraphics[width=1\linewidth]{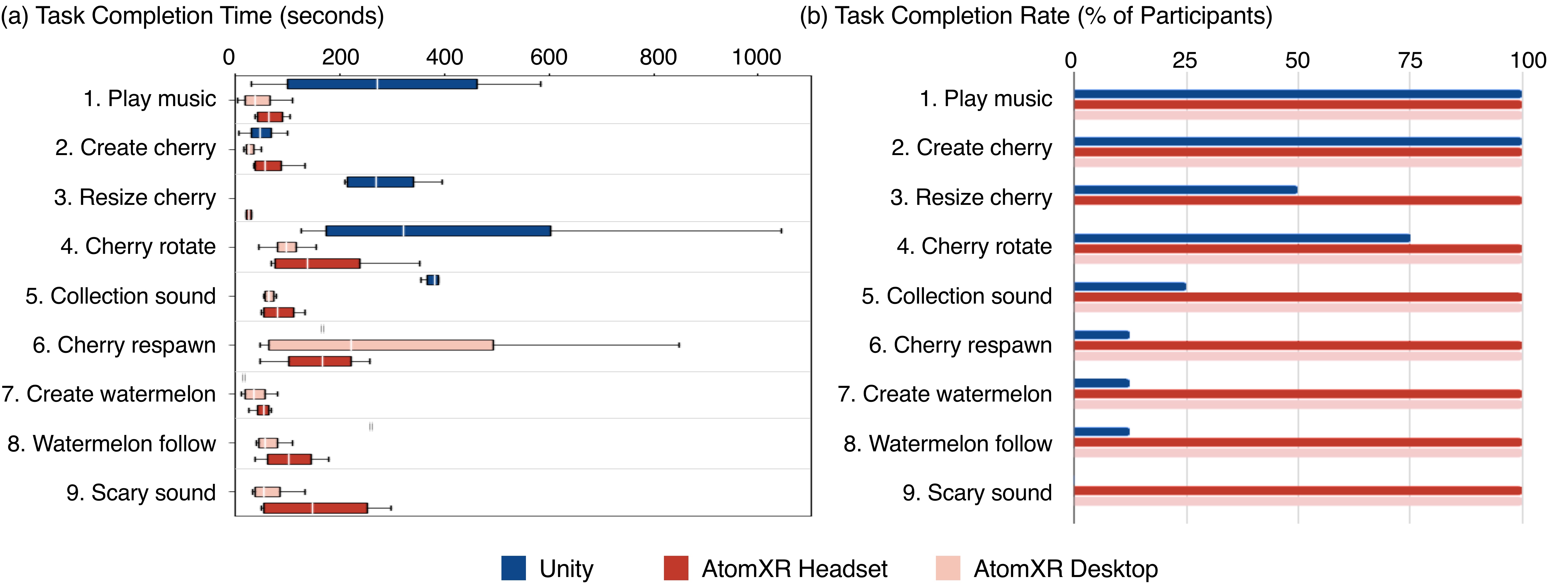}
\vspace{-4mm}
\caption{Study \#1 results showing (a) task completion time and (b) task completion rate for three conditions (Unity, AtomXR Desktop, and AtomXR Headset) across all tasks \protect\footnotemark.}
\vspace{-4mm}
\label{fig:study1-tasktime}
\Description{Box and whisker plots of task completion times in three experimental conditions (Unity, AtomXR Desktop, AtomXR Headset) for nine different tasks. The tasks include 1) play music, 2) create cherry, 3) resize cherry, 4) cherry rotate, 5) collection sound, 6) cherry respawn, 7) create watermelon, 8) watermelon chase, 9) scary sound, and are further are detailed in Table 1. The x axis shows each task and the y axis shows the task completion time in seconds. The Unity task completion times generally show a larger interquartile range than the AtomXR Headset and AtomXR desktop task completion times.}
\end{figure}
\footnotetext{When interpreting task-by-task graphs, note that tasks were given in sequential order, so if a participant was stuck on one task, they would be unable to complete subsequent tasks. The data shown in task completion graphs are from participants who completed the task, so time variances may not be directly compared across tasks.}

\subsection{User Performance}

In the AtomXR Headset and AtomXR Desktop conditions, all 8 participants successfully completed 100\% of the tasks, with an average of 6.4 minutes to spare. By contrast, none of the participants were able to complete all tasks in the Unity control condition, and on average were only able to complete 46\% of tasks. Moreover, the average task completion time was 98.25 seconds with AtomXR Headset, 93.34 seconds with AtomXR Desktop, and 243.49 seconds with Unity.

Fig.~\ref{fig:study1-tasktime} shows the average task completion times across tasks, and Fig.~\ref{fig:study1-experience}a shows the percentage of participants able to complete each task.\footnote{\label{note1}Task~\#3 was ommitted for the AtomXR Desktop condition due to the inability to use holographic remoting with AtomXR Desktop.} From this breakdown, we observe that tasks requiring implementation of logic or manipulation of 3D objects are performed much faster using AtomXR systems compared to Unity.\footnote{One participant ran into a bug in Task~\#6, resulting in anomalous data for AtomXR Desktop in that task.} On the other hand, simple object creation tasks (e.g., Task~\#2 for creating the cherry and Task~\#7 for creating the watermelon) are faster with Unity's drag-and-drop interface. However, in workflows in which users need to search online to find models outside of built-in models, AtomXR's automated search process that queries models from an online database based on semantic similarity to the user's request may provide value. 


\subsection{User Experience \& Preference}

\begin{figure}[t]
\centering
\includegraphics[width=.55\linewidth]{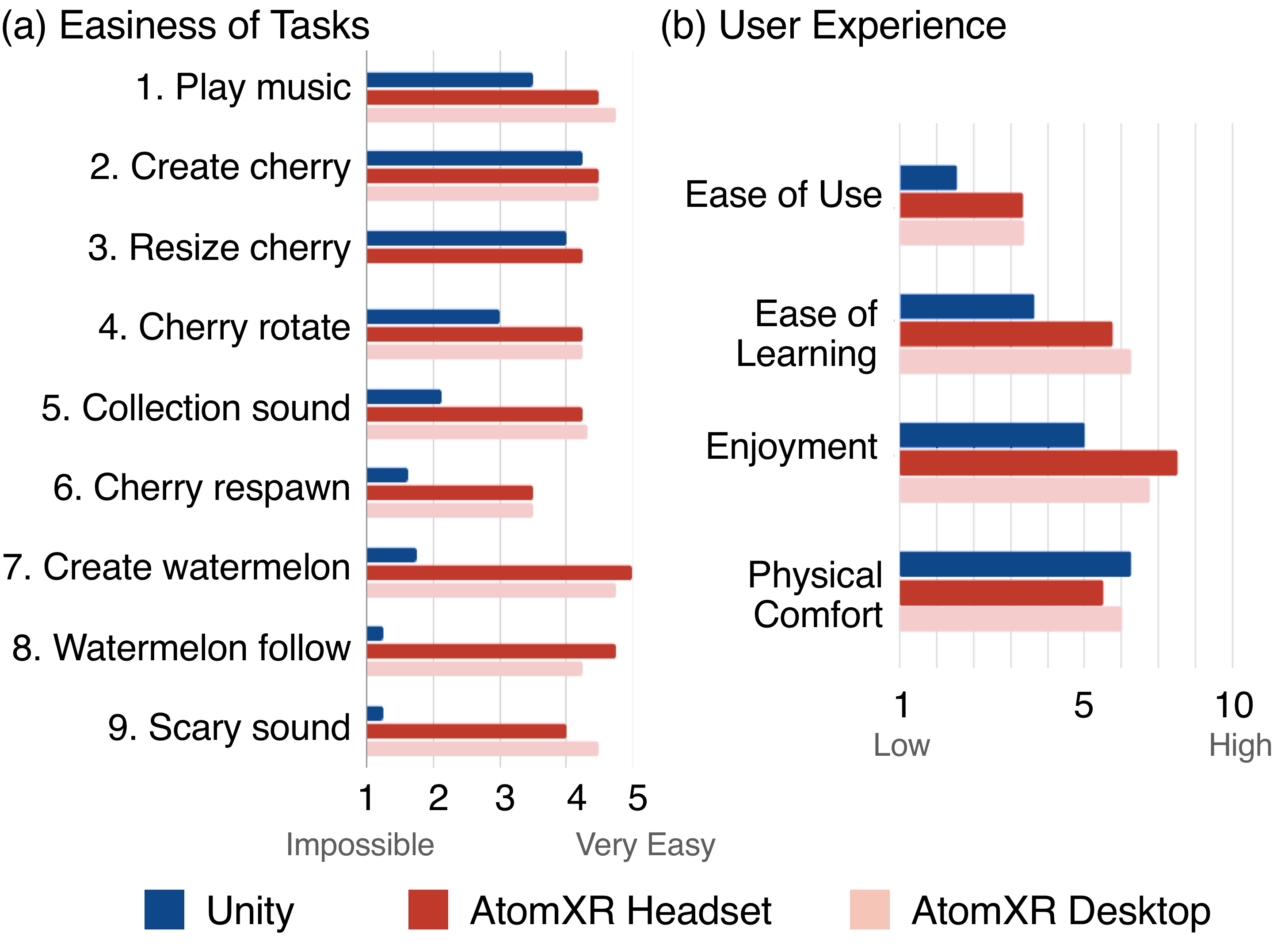}
\vspace{-4mm}
\caption{Study \#1 user ratings comparing Unity, AtomXR Headset, and AtomXR Desktop in the easiness of tasks and user experiences.}
\label{fig:study1-experience}
\Description{Bar graph showing the participant-reported evaluations of the overall experience, ease of use, ease of learning, enjoyment, and physical comfort across three experimental conditions (Unity, AtomXR Desktop, AtomXR Headset). AtomXR Headset and AtomXR Desktop received higher evaluations than Unity in all but the physical comfort category.}
\end{figure}

Beyond objective performance improvements, users also experienced the tasks as being easier to implement using AtomXR compared to Unity, as shown in Fig.~\ref{fig:study1-experience}a. Reflecting on their overall experience with both Unity and AtomXR systems, participants found AtomXR systems easier to use and easier to learn, and also found the experience more enjoyable, albeit slightly less physically comfortable, as depicted in Fig.~\ref{fig:study1-experience}b. Some participants felt intimidated by Unity, with P6 noting that Unity \textit{"trigger[ed] [in] me the fear of learning coding"}, and others felt some uncertainty with AtomXR, noting they \textit{"feel less confident about troubleshooting mistakes while prototyping"} (P1).

In terms of user preferences, participants generally felt that the relative utility of each system depends on the context, with P8 noting that \textit{"there's an accessibility/extensibility trade-off"}. Participants often found AtomXR \textit{"fun and easy"} (P4), and noted that they \textit{"would like to use AtomXR to get prototype and getting start with design games, but Unity for further development"} (P6).

Based on these results, we conducted a follow-up study to further explore the design implications of each component (scripting abstraction, natural language based scripting, immersive environment).

\section{Study \#2: Components of Multimodal Prototyping Experience}

This study aimed to dive into each component of the system and investigate its impact on user performance and experience in order to better understand design opportunities for the future of XR development. Specifically, this study explored the following research questions:

\begin{itemize}
    \item \textbf{RQ3:} \emph{How do users interact with and value each feature of AtomXR, including AtomScript, natural language interaction, and immersive authoring?}
    \item \textbf{RQ4:} \emph{How do variations of the system with different features compare in terms of learnability, ease of use, and efficiency for prototyping interactive XR applications?}
\end{itemize}

\subsection{Study Design}

Similar to the first study, the main study was also a hybrid within-subjects and across-subjects study, in which participants again were given 20 minutes to complete a series of development tasks using two different development systems. In this study, however, we focused on isolating the interaction dynamics and contributions of each feature of the AtomXR system. For this purpose, we designed four conditions with different combinations of abstraction, natural language scripting, and immersive authoring:
\begin{itemize}
    \item \emph{Unity with C\#} - Participants use the traditional Unity development system with their standard C\# scripting system.
    \item \emph{AtomScript} - Participants use the traditional Unity development environment but are able to script using the high-level AtomScript language.
    \item \emph{AtomXR Desktop} - Participants use the AtomXR system in a desktop environment, in which they build applications on a 2D screen by giving natural language requests.
    \item \emph{AtomXR Headset} -  Participants use the AtomXR system in a headset environment, in which they can experience and edit applications in a 3D AR environment using physical manipulation, eye gaze-assisted object referencing, and natural language requests.
\end{itemize}

We designed four treatments, each containing two of these four conditions, to investigate the effect of each new feature. Four participants experienced each treatment, and the ordering of conditions within treatments was counterbalanced to minimize any learning effect. The treatments were as follows:
\begin{itemize}
    \item \emph{Unity with C\# vs Unity with AtomScript} - This comparison investigates the effect of allowing users to create logic using a high-level programming language that abstracts away the low-level complexities of the standard C\# system.
    \item \emph{AtomScript vs AtomXR Desktop} - This comparison investigates the effect of allowing users to create logic and objects using natural language instead of writing code and using a graphical user interface. 
    \item \emph{AtomXR Desktop vs AtomXR Headset} - This comparison investigates the effect of allowing users to author in an immersive, 3D environment instead of on a 2D interface.
    \item \emph{AtomXR Headset vs Unity} -  This comparison investigates the integrative effects of all features of AtomXR. 
\end{itemize}

\subsection{Participants \& Procedure}

We recruited 16 participants (4 female, 12 male) from university and public interest group channels, selecting for varied levels of prior experiences and perspectives to create a balanced pool. On average, participants held positive attitudes toward new technologies (mean = 4.81, SD = 0.40). On a scale from 1 (layman) to 5 (expert), participants on average reported a 3.63 level of prior programming experience (SD = 1.20), a 2.19 level of prior game development experience (SD = 1.11), and a 1.75 level of prior experience working with AR/VR (SD = 1.00). 

The procedure for this study followed the same structure as the procedure for the first study, with the only differences being in the systems participants used and the questionnaires they were given. In this study, the surveys given were more in-depth, and included open-ended questions asking about user experiences with each component of the system (high-level language, natural language logic creation, immersive authoring, etc.).

\section{Results}

Combining the quantitative analyses and ratings with qualitative responses from participants’ open-ended answers and interviews, we synthesized the following insights on user experiences, perception, and interaction with different features of the system.
\begin{figure}[t!]
\centering
\includegraphics[width=\linewidth]{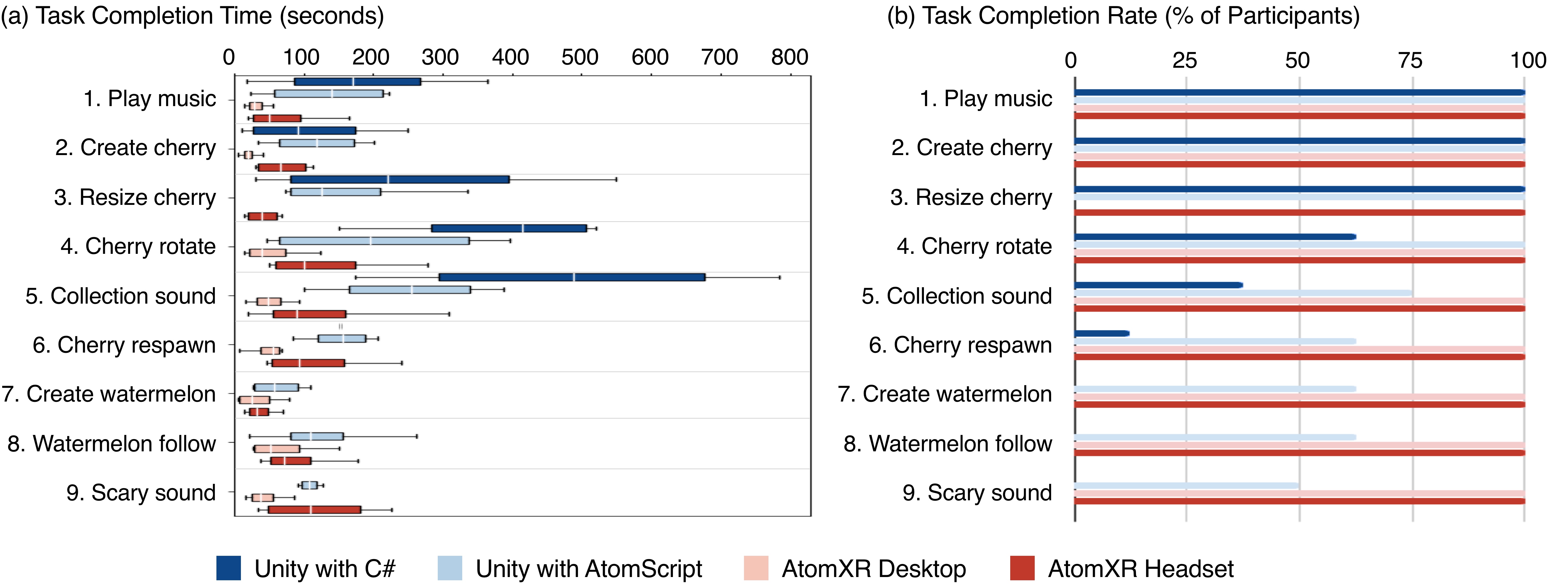}
\vspace{-4mm}
\caption{Study \#2 results showing (a) task completion time and (b) task completion rate for four experimental 
 conditions (Unity with C\#, Unity with AtomScript, AtomXR Desktop, and AtomXR Headset) across all tasks.}
 \vspace{-4mm}
\label{fig:study2-results}
\Description{Box and whisker plots of task completion times in four experimental conditions (Unity with C#, Unity with AtomScript, AtomXR Desktop, AtomXR Headset) for nine different tasks. The tasks are detailed in Table 1. The x axis shows each task and the y axis shows the task completion time in seconds. The Unity with C# task completion times generally show a higher average and larger interquartile range than the Unity with AtomScript, AtomXR Headset, and AtomXR desktop task completion times.}
\end{figure}

\subsection{AtomScript enables improved task completion and development experience compared to traditional C\# (Unity with C\# vs. Unity with AtomScript)}

Fig.~\ref{fig:study2-results} shows the average task completion times across tasks (Fig.~\ref{fig:study2-results}a) and the percentage of participants able to complete each task (Fig.~\ref{fig:study2-results}b).
Within the allotted time frame, participants were able to complete on average 45\% of the tasks in the \textit{Unity with C\#} condition and 79\% of tasks in \textit{Unity with AtomScript}. They completed tasks on average 1.48 times as fast using \textit{Unity with AtomScript} compared to \textit{Unity with C\#}, suggesting a significant barrier and point of friction to early stage application prototyping is the complexity of using low-level C\# code.

 From Fig.~\ref{fig:study2-results}a, we observe that \textit{Unity with AtomScript} provides general speed improvements over \textit{Unity with C\#} for all available tasks except for the object creation task (i.e., Task~\#2 for creating the cherry). In particular, AtomScript provided a large speed improvement in Task~\#4 (make the cherry rotate) and Task~\#5 (make the cherry play a collection sound when the player runs into it). This could potentially be explained by the large reductions in syntax and implementation complexity of rotation and collision detection logic from C\# to AtomScript. This highlights the potential advantages of high-level functional abstractions in accelerating and simplifying early-stage prototyping for users. Specifically, AtomScript offers significant benefits in scenarios where traditional low-level implementations are syntactically complex or require multiple steps.

\begin{figure}
\centering
\includegraphics[width=1\linewidth]{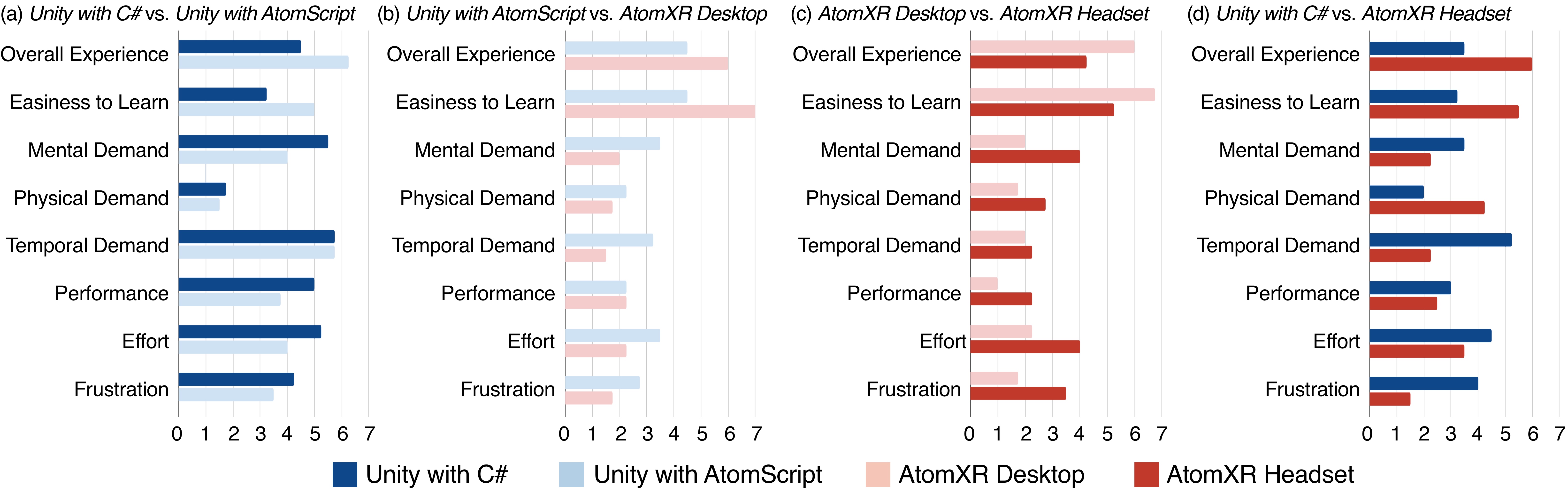}
\caption{Study \#2 user ratings comparing four treatment groups across experience metrics from the NASA TLX set, along with two additional evaluations of overall experience and easiness to learn, which are reversed in polarity in the graph.}
\Description{Study #2 user ratings comparing four treatment groups across experience metrics from the NASA TLX set, along with two additional evaluations of overall experience and easiness to learn, which are reversed in polarity in the graph. The data shows a single bar chart for each of the conditions, where a given bar chart has colors differentiating between two of each of the 4 treatment types.}
\vspace{-4mm}
\label{fig:study2-experience}
\end{figure}

Comparing the two conditions, we also found experience improvements in ease of learning, overall user experience, and nearly all dimensions of evaluation measured by the NASA TLX set \cite{hart2006nasa}, as shown in Fig.~\ref{fig:study2-experience}a. The most notable improvements were in the overall experience and the easiness to learn the system (39\% and 54\% improvement of \textit{Unity with AtomScript} over \textit{Unity with C\#}, respectively). All four participants in this treatment found \textit{Unity with AtomScript} to be easier to use, and preferred it over \textit{Unity with C\#} for prototyping. Three out of four participants found \textit{Unity with AtomScript} easier to learn, while one participant thought \textit{Unity with C\#} was easier because there was more online documentation that could be copied and pasted in contrast to AtomScript (P2). 

In general, participants appreciated the reduction in complexity of the code they needed to write, with P2 observing, AtomScript \textit{"allowed me to significantly reduce the amount of code I had to physically type out"} and \textit{"reduced the number of errors that arose during compilation/runtime"} and P4 saying that the \textit{"complicated syntax of C\# made it difficult to pick up and use"} in comparison to AtomScript. Regarding the impact of AtomScript on developer experiences, P6 noted that \textit{"reduced stress level leads to more creativity"}. In general, when discussing the trade-offs of AtomScript and C\#, participants noted that AtomScript lacked the community support that an established language like C\# had, while C\# lacked beginner friendliness and would at times be over complicated for the purpose it served.


\subsection{Natural language accelerates the development process and enables more creativity (Unity with AtomScript vs. AtomXR Desktop)}

The introduction of natural language driven AtomScript generation allowed users to complete 100\% of tasks (Fig.~\ref{fig:study2-results}b), and complete tasks far faster than by directly writing AtomScript. On average, participants completed tasks 3.38 times as fast using \textit{AtomXR Desktop} compared to \textit{Unity with AtomScript} (Fig.~\ref{fig:study2-results}a).

In general, participants noted several components of the \textit{AtomXR Desktop} system they appreciated, particularly the use of natural language and the reduction of context switching. Reflecting on the two systems, P12, an expert in prior programming experience, said, \textit{"Natural language is a lot easier than code, even for a relatively experienced coder"}, and other participants similarly expressed that the ability to use natural language made the tasks easier and more stress-free. Beyond natural language, P13 observed that \textit{"one big advantage of the AtomXR userface is that it is obviously not as bloated, so switching from Edit to Test only takes a second - whereas it takes much longer to test anything in the Unity environment"}. Moreover, \textit{"AtomXR has a huge advantage of having a naturally aligned play and edit mode: mapping the axes on the monitor [in Unity] to real life 3D was nonobvious and challenging"} (P12). In contrast to AtomXR, Unity's fragmentation between play and edit mode, combined with the need to change software to find assets and write logic (P13 found the \textit{"switching between VSC \footnote{VSC = Visual Studio Code} and Unity [to be] annoying"}), creates constant context switching that may feel overwhelming, particularly for beginners. Avoiding this cognitive load, participants found the ability to even be more creative, with P13 saying, \textit{"AtomXR saves so much time and I feel like it gives me creative superpowers"}. However, participants noted system features that could use improvement, with P7 saying they \textit{"wish AtomXR gave the ability to manually edit the code, or look at text prompt history, since that would greatly increase developer workflow"} and P13 noting that \textit{"AtomXR is still lacking some basic UI improvements"}. These limitations are addressed in Sec.~\ref{sec:limitations}.

One interesting observation based on Fig.~\ref{fig:study2-results}a is that there exist significant differences in variance of average task completion time across tasks for each system. In particular, the variance for \textit{Unity with C\#} is approximately 21502 seconds, which is reduced around ten-fold to around 2988 for \textit{Unity with AtomScript}, and ten-fold again to around 208 for \textit{AtomXR Desktop}. This is likely because natural language input takes similar amounts of time for commands of different levels of complexity, while time taken to write code scales with complexity the more low-level the programming language is. 


\subsection{Immersion slows down the process due to input method issues but enhances fine-tuning objects in space (AtomXR Desktop vs. AtomXR Headset)}

Interestingly, the addition of the immersive authoring component resulted in worse quantitative and qualitative metrics overall in the \textit{AtomXR Headset} condition in comparison to \textit{AtomXR Desktop}. Participants completed tasks on average 0.57 times as fast using \textit{AtomXR Headset} compared to \textit{AtomXR Desktop}. Still, \textit{AtomXR Headset} allowed users to complete 100\% of tasks in the allocated time.


In terms of user experience, \textit{AtomXR Desktop} performed better along all dimensions, as shown in Fig.~\ref{fig:study2-experience}c. Qualitative data speaks to the main reason why participants felt this way. P14 explained that their headset performance was worse because of \textit{"lag in pressing buttons and dictation"} and \textit{"having to go back and edit text that didn't come through"} in dictation. Issues with pressing buttons, dictation inaccuracies, and other difficulties interacting with the 3D in-headset interfaces, caused the development system to be harder to use and slower. Participants tended to prefer familiar desktop input methods, as \textit{"being able to use the mouse and type with a keyboard was much faster and less physically strenuous than headset"} (P14). Because of this, all four participants who experienced the desktop and headset variations preferred the desktop variation for early prototyping and found it easier to use. 

Participants did, however, credit the immersive authoring environment for making it easier to move around, and visualize and manipulate objects. This is reflected quantitatively, as \textit{AtomXR Headset} provides a significant improvement in speed for tasks involving editing of objects to match end-user perception (i.e., Task~\#3 for resizing the cherry to be life-sized). In the non-immersive authoring conditions, participants had to change the size using a desktop interface, stream the application to the headset for testing, and then go back and adjust the size again if their original estimation was incorrect. Being able to use hands to directly manipulate the object in 3D was a clear improvement for these types of authoring tasks. Moreover, participants appreciated other immersive methods, such as the \textit{"eye-based "this" feature for naming objects instead of using their object id name"} (P10). Based on each system's advantages, P14 suggested \textit{"combining the two systems, and using the VR for its advantages - spatial visualization, tactile manipulation - and using the desktop for its advantages of speed, precision of language input, and lack of physical exertion"}. 


\subsection{Immersive authoring with natural language enhances overall development experience (Unity with C\# vs. AtomXR Headset)}

Finally, comparing \textit{Unity with C\#} with \textit{AtomXR Headset}, we found participants completed tasks on average 2.84 times as fast using \textit{AtomXR Headset} and were able to complete 100\% of the tasks using \textit{AtomXR Headset} in contrast to only 45\% using \textit{Unity with C\#}. We also found experience improvements along all dimensions of evaluation except physical demand, which is reasonably explained by the strain of wearing the headset. Specifically, participants reported a 72\% improvement in overall experience and 69\% improvement in learning ease using \textit{AtomXR Headset} over \textit{Unity with C\#}. Impressively, the average frustration experienced by participants was reduced by 63\% using \textit{AtomXR Headset}, followed by a 57\% reduction in temporal demand, 36\% reduction in mental demand, and 22\% reduction in effort. Participants unanimously preferred AtomXR over Unity for early prototyping, and found it easier to use and easier to learn, supporting the results from Study \#1. 

Beyond being \textit{"simple to use"} and \textit{"alot easier to use for prototyping"} (P9), participants also \textit{"felt great after [using AtomXR]"} (P8) and found it \textit{"more fun to get results immediately after saying something"} (P9). Consistent with previous reports from participants, P1 explained that the ability to use natural language \textit{"made the tasks so much easier, [as they] did not have to worry about learning the syntax or other programming related details"}. P9 elaborated that \textit{"using natural language helped with the performance because it is a more direct connection between your thoughts and input."} However, participants expressed concerns about the customizability and controllability of AtomXR, noting that \textit{"it seems like [Unity] provides the developer with way more room for customization"} (P1).

Reflecting on in-headset authoring, participants again expressed it was \textit{"much easier to get a sense of space"} (P5), which made manipulating the size and position of objects easier. Beyond functionality, the immersion of the authoring environment made users enjoy the experience more and perceive it as fun, rather than as purely a task- \textit{"Once I got the hang of it, I kind of enjoyed it, and played it like a game,"} P8 commented. However, participants again reported frustrations with in-headset issues like lag time in interactions, interface sensitivity, and other issues with the HoloLens hardware.

\begin{figure}
\centering
\includegraphics[width=.55\linewidth]{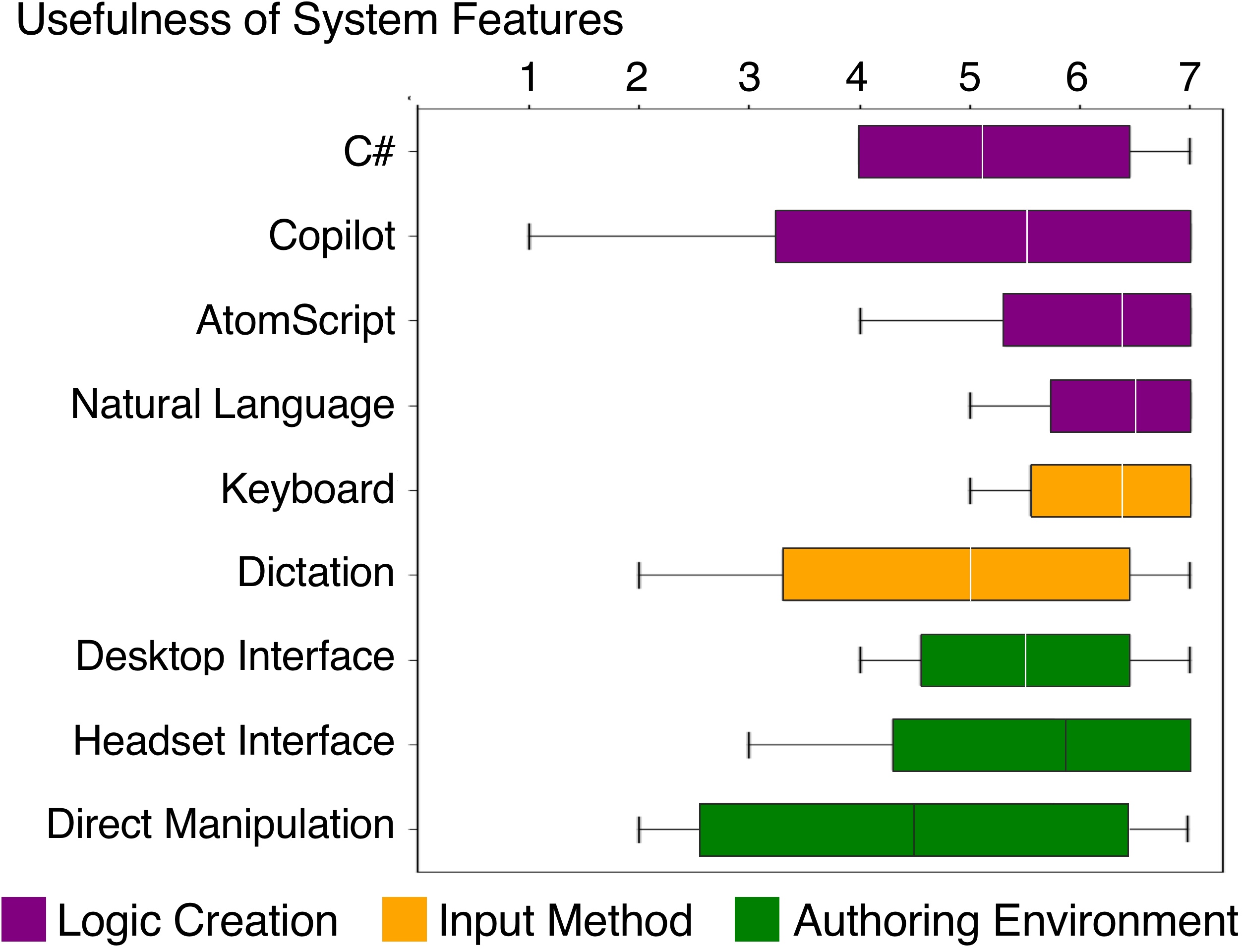}
\vspace{-1mm}
\caption{User ratings on the usefulness of each component of the system, including approaches to logic creation, input methods, and authoring environments.}
\label{fig:study2-features}
\Description{Box and whisker plots of the usefulness of 9 different features of different development systems. Specifically, the x axis shows the features, including logic creation features (C#, Copilot, AtomScript, natural language), input method features (keyboard, dictation), and authoring environment features (desktop interface, headset interface, and direct manipulation). The y axis shows the average participant rating, from 1 (useless) to 7 (very useful). The features with the highest average rating were natural language, AtomScript, and the keyboard.}
\end{figure}
\vfill

\textbf{Usefulness of Features.}
Aggregating the usefulness ratings for all different features from both the traditional Unity system and AtomXR Headset in Fig.~\ref{fig:study2-features}, we found that the major trade-off made between the traditional Unity system and the full AtomXR Headset system revolves around the affordances of the device hardware. In developing the system, we had hoped that dictation would be a more natural form of interaction---however, participants indicated in their usefulness ratings and open-ended comments that they preferred the traditional keyboard as the input method within development processes. Additionally, although participants appreciated the ability to directly manipulate objects from the perspective of their end user, the benefits of this feature could not outweigh the slowness of headset interaction. While in-headset input methods are expected to improve in the future, considering these trade-offs, P5 suggested \textit{"AtomXR could use a hybrid system where I could do some authoring on the desktop."}

Interestingly, when considering the integrated AtomXR Headset system, participants tended to notice the natural language and immersive authoring features more than AtomScript. This may be because the tasks in the experiment didn't necessitate extensive debugging or customization, reducing the need for participants to engage deeply with the generated AtomScript code.
This illuminates the importance of different features based on application context, suggesting initial logic generation through natural language may be crucial during early prototyping stages, while fine-tuning via AtomScript becomes more vital as the project matures or becomes more complex. 
These two components of the system complement and depend on each other---the more robust the natural language code generation, the less reliant users become on AtomScript for adjustments (we examine the current limitations of our NLP methods in Appendix ~\ref{sec:limitations}.

\section{Discussion and Future Work}
\label{sec:discussion}


\textbf{Reduced intent-to-input translation contributes to better development experience.} Our results indicate that our immersive natural language driven authoring system is generally easier to learn and use than the traditional Unity development system. While the abstraction provided by AtomScript contributed to this ease, participants found the most added value in the ability to use natural language to describe their requests. Especially in logic creation tasks, participants appreciated the direct translation from intent to system input within AtomXR in comparison to the longer process required to translate intent to code using traditional methods. This lower cognitive demand often improved participants' sense of enjoyment during the development process. In object creation tasks, however, using natural language as an input method often slowed participants down. This can be explained by the innate difference between how people imagine different tasks--language is more often used in imagining functionality than in imagining visual designs. Thus, it may be more intuitive to use visual methods such as drag-and-drop GUIs for object or world creation tasks. Based on these insights, we believe design of future development tools would benefit from designing affordances based on characterization of how the user imagines tasks anticipated in the authoring process.

\textbf{Immersive authoring is not yet preferred due to inconvenient input methods.} The in-headset variation allowed participants to get a better sense of their end user's experience, reducing the gap between the development and testing process, in particular for manipulating objects in 3D space. The immersive, game-like environment of the in-headset variation was also found to be more fun and engaging. However, these advantages were outweighed by the difficulties of current input and interaction methods for headset devices. The frustrations expressed regarding lag and inaccuracies in dictation and touch interactions suggest that hardware issues still may present a barrier to intensive usage of immersive authoring systems. 

Based on these insights, we believe design of future development tools should optimize usability based on the provided affordances--given difficulties with direct text input and editing, systems can develop methods of communication focused less on precision. More modalities of input, such as eye gaze and gesture as we see in AtomXR, and better understanding of user-system interaction history can help elucidate imprecise requests and minimize the impact of input issues.


\textbf{Heading toward streamlined and naturalized prototyping environments.} Beyond natural language, participants appreciated AtomXR's streamlined environment, in which asset selection, world design, logic creation, and testing all happened within the same environment, eliminating the cognitive and time cost of context switching necessary in the traditional desktop systems. Most participants performed better and preferred using all variations of the AtomXR authoring system over traditional methods. The trend toward streamlined development is reflected in industry as well, where NVIDIA's Omniverse \cite{omniverse} attempts to provide a cohesive platform for development, deployment, and management of 3D applications. Based on our results, we believe that streamlined and naturalized authoring environments provide significant value to early stages of the XR prototyping process and make XR more accessible to a wide range of creators from various fields. 


\textbf{Limitations.}
Our studies and system have several limitations. The studies focused on a set of relatively simple development tasks. To fully gauge AtomXR's utility in real-world settings, a broader array of more complex usage scenarios should be evaluated. Additionally, the studies compare AtomXR with Unity to understand relative advantages and disadvantages of our system and a commonly-used traditional development system, but further studies are necessary to understand AtomXR's utility relative to other development tools.

%
In terms of system limitations, AtomXR is currently a proof-of-concept for a natural language-driven immersive prototyping tool. As such, it is not as capable as full development systems like Unity. For example, the current implementation of AtomXR does not allow users to build and deploy their applications to platforms outside of HoloLens 2. As suggested by study participants, AtomXR could be integrated with existing authoring workflows. AtomXR and Unity could complement each other in the development pipeline, where the developer can quickly create 3D worlds and prototype basic logic in AtomXR and then use Unity to add more complex functionality such as integrating 3rd party libraries or creating networking requests.

Another challenge lies in the system's accuracy in converting natural language intent to valid AtomScript code. GPT-3 often generates functions that are not implemented in AtomScript. See Appendix ~\ref{sec:limitations} for examples of errors in natural language to AtomScript translation. Specifically, Yin and Neubig \cite{yin-neubig-2017-syntactic} have shown that it’s possible to parse natural language descriptions into syntactically valid code using a novel neural architecture. Future work could use this model to take in the underlying grammar of AtomScript as prior knowledge to ensure that generated AtomScript is correct. 

Beyond initial generation accuracy, AtomXR's editing affordances are sparse. Editing in immersive authoring presents a major challenge due to the difficulty of text input in XR ~\cite{kern2023text, yang2022flick}. A potential solution to explore in future work is visual scripting as a method for fine-tuning generated logic. AtomXR could also benefit from retaining and using a history of user interactions as context for understanding user intents. Being a research prototype, AtomXR also lacks the support, documentation, and community of more popular tools. Moreover, the user interface could be dramatically improved in future work.  



\textbf{Implications for Future XR Prototyping.}
In the evolving landscape of XR and AI-driven content, our work reflects the shift from passive observation to active participation in world creation. By making functional XR development accessible, we blur the lines between participant and author towards shared, peer-to-peer realities. Our use of NLP techniques to power our system reflects the larger rise in usage of NLP technologies across fields, which may have fundamental implications for the ways in which we think and create in the future.

\section{Conclusion}
In this paper, we introduce AtomXR, an innovative XR authoring system that leverages NLP with multimodal interactions to simplify the prototyping process. We contributed a streamlined immersive authoring environment, natural language driven multimodal interaction scheme, and a custom, high-level language designed for XR prototyping. In a two-part user study, we evaluated AtomXR on typical XR development tasks and found it not only accelerated task completion by 2-5 times but was also more intuitive and easier to learn. We further compared the effects of immersion, natural language, and AtomScript on development experiences. Based on these findings, we suggest avenues for future research to refine this promising approach to XR prototyping.

\bibliographystyle{ACM-Reference-Format}
  \bibliography{AtomXR}

\section{Appendices}

\subsection{Survey Questions}

The following are the post-task survey questions participants answered after each task:
\begin{itemize}
    \item How easy was it to complete each task with the system? (rating)
    \item How would you rate your overall experience with this system? (rating)
    \item How easy was it to learn how to use this system? (rating)
    \item How would you rate your physical comfort while using this system (e.g. eye strain, head ache, etc.)? (rating)
    \item How mentally challenging was the task? (rating)
    \item How physically challenging was the task? (rating)
    \item How hurried or rushed was the pace of the task? (rating)
    \item How successful were you in accomplishing what you were asked to do? (rating)
    \item How hard did you have to work to accomplish your level of performance? (rating)
    \item How insecure, discouraged, irritated, stressed, and annoyed were you? (rating)
\end{itemize}

The following are the post-experiment survey questions participants answered after completing the entire experiment:
\begin{itemize}
    \item Which system was easier to use? (multiple-choice)
    \item Which system was easier to learn? (multiple-choice)
    \item Which system would you prefer for early prototyping? (multiple-choice)
    \item Which system did you prefer overall and why? (open-ended)
    \item What other functionality do you think each system is lacking? (open-ended)
    \item What did you like most about each tool? (open-ended)
    \item What did you dislike most about each tool? (open-ended)
    \item Reflect on the differences in your experience using each system. What did you like and dislike about each? (open-ended)
    \item How useful was each of the features in assisting with development? (rating)
    \item Describe if and how each feature impacted your performance on the task in comparison to using Unity. (open-ended)
\end{itemize}

\subsection{AtomScript Specification}
\label{sec:atomscriptgrammar}

For reproducibility, the grammar of AtomScript is defined below in G4 file format (used to define grammars for ATNLR4).

\begin{lstlisting}[style=g4style]
grammar Hello;

program: line* EOF;

line: statement | ifBlock | foreverBlock | onStartBlock | onCollisionBlock | onButtonPressBlock;

statement: (assignment | functionCall) ';';

ifBlock: 'if' expression block ('else' elseIfBlock)?;

elseIfBlock: block | ifBlock;

foreverBlock: FOREVER block;

FOREVER: 'forever';

onStartBlock: ONSTART block;

ONSTART: 'onStart';

onCollisionBlock: ONCOLLISION '<' constant ',' constant '>' block;

ONCOLLISION: 'onCollision';

onButtonPressBlock: ONBUTTONPRESS '<' constant '>' block;

ONBUTTONPRESS: 'onButtonPress';

COMMENT: '/*' .*? '*/';

LINE_COMMENT: '//' ~[\r\n]*;

assignment: IDENTIFIER '=' expression;

functionCall:
	IDENTIFIER '(' (expression (',' expression)*)? ')';

array
 : '[' ( expression ( ',' expression )* )? ']'
 ;

expression:
	constant
	| IDENTIFIER
	| array
	| functionCall
	| '(' expression ')'
	| '!' expression
	| expression multOp expression
	| expression addOp expression
	| expression compareOp expression
	| expression boolOp expression;

multOp: '*' | '/';
addOp: '+' | '-';
compareOp: '==' | '!=' | '<' | '>' | '<=' | '>=';
boolOp: '&&' | '||';

INTEGER: [0-9]+;
FLOAT: [0-9]+ '.' [0-9]+;
STRING: ('"' ~'"'* '"') | ('\'' ~'\''* '\'');
BOOL: 'true' | 'false';
NULL: 'null';

constant: INTEGER | FLOAT | STRING | BOOL | NULL;

block: '{' line* '}';

WS: [ \t\r\n]+ -> skip;
IDENTIFIER: [a-zA-Z_][a-zA-Z0-9_]*;
\end{lstlisting}

\subsection{Natural Language to AtomScript Translation Error Examples}
\label{sec:limitations}
We present several examples of natural language-code pairs and compare the code outputted by GPT-3 with the expected code output to demonstrate the limitations of the current NLP process (see Table \ref{tab:nlp}).

We observed that in examples 1, 2, and 5, the AtomScript generated by GPT-3 worked equivalent to the reference AtomScript. However, in examples 3 and 4, the generated AtomScript showed two common failure points. 

Firstly, in example 3, GPT-3 did not define the scoreboard variable beforehand, likely because the boundaries of this natural language request are unclear. This kind of request could often appear in a context in which the user has already defined a scoreboard system. GPT-3 also omits the implementation of the collision logic, likely because it assumes these functions have been implemented in surrounding context. To address this, we plan to store a history of past commands to give more context.

The other failure point was in example 4, where GPT-3 referenced a function that actually doesn't exist in our system (there is no "Wait" function). Although this could be easily implemented and is a plan for future work, a syntax-based language model would be a better approach for generating valid code.


\begin{table*}
\centering
\caption{\label{tab:nlp}: A comparison of AI generated AtomScript translation and reference translation}
\begin{tabular}{|p{.2\textwidth}|p{.39\textwidth}|p{.39\textwidth}|}\hline
Prompt  &  GPT Generated AtomScript & Reference AtomScript \\
\hline
Whenever the player collects coin1, play a coin collection sound&\begin{lstlisting}[aboveskip=-6pt, belowskip=-1 \baselineskip]
onCollision<"Player", "coin1"> {
    PlaySound("Coin collect");
}
\end{lstlisting}&\begin{lstlisting}[aboveskip=-6pt, belowskip=-1 \baselineskip]
onCollision<"Player", "coin1"> {
    PlaySound("Coin collect");
}
\end{lstlisting}\\\hline
Make box1 disappear if at least 5 seconds have passed and I’ve hit it twice & \begin{lstlisting}[aboveskip=-6pt, belowskip=-1 \baselineskip]
currTime=TimeSinceStart();
numCollisions=0;
onCollision<'Player','box1'> {
    numCollisions=numCollisions+1;
}
forever {
    if(TimeSinceStart()-currTime >= 5 && numCollisions>=2) {
        Disappear('box1');
    }
}
\end{lstlisting} & \begin{lstlisting}[aboveskip=-6pt, belowskip=-1 \baselineskip]
numCollisions = 0;
forever {
	if (TimeSinceStart() > 5) && (numCollisions >= 2) {
		Disappear("box1");
	}
}
onCollision<"Player", "box1"> {
	numCollisions = numCollisions + 1;
}
\end{lstlisting}\\\hline

Make it so that enemy1 starts following me after I get 5 coins or 30 seconds have passed & \begin{lstlisting}[aboveskip=-6pt, belowskip=-1 \baselineskip]
forever {
    if(scoreboard>=5 || TimeSinceStart()>=30) {
        Move('enemy1','slow',GetPosition('Player')-GetPosition('enemy1'));
    }
}
\end{lstlisting} & \begin{lstlisting}[aboveskip=-6pt, belowskip=-1 \baselineskip]
coinsCollected = 0;

forever {
    if (coinsCollected >= 5) || (TimeSinceStart() > 30) {
        Move("enemy1", "slow",  GetPosition("Player") - GetPosition("enemy1"));
    }
}

onCollision<"Player", "Coin1"> {
    coinsCollected = 1 + coinsCollected;
}
\end{lstlisting}\\\hline

Make apple1 flash on and off every second & \begin{lstlisting}[aboveskip=-6pt, belowskip=-1 \baselineskip]
forever {
    ChangeColor('apple1',[1,0,0]);
    Wait(1);
    ChangeColor('apple1',[0,0,0]);
    Wait(1);
}
\end{lstlisting} & \begin{lstlisting}[aboveskip=-6pt, belowskip=-1 \baselineskip]
isOn = true;
timeElapsed = 0;
iter = 0;
forever {
  if(timeElapsed >= 1){
		if(isOn){
			Disappear("apple1");
			isOn = false;
		}
		if(isOn == false){
			Appear("apple1");
			isOn = true;
		}
		iter = iter + 1;
	}
	timeElapsed = timeSinceStart() - iter*1;
}
\end{lstlisting}\\\hline

Make it so that cube2 chases me. & \begin{lstlisting}[aboveskip=-6pt, belowskip=-1 \baselineskip]
forever {
    Move('cube2','fast',GetPosition('Player')-GetPosition('cube2'));
}
\end{lstlisting} & \begin{lstlisting}[aboveskip=-6pt, belowskip=-1 \baselineskip]
forever{
    Move('cube2','fast',GetPosition('Player')-GetPosition('cube2'));
}
\end{lstlisting}\\\hline
\end{tabular} 
\Description{Three column table demonstrating the limitations of the NLP system, containing the Prompt. the Generated AtomScript, and the Reference AtomScript. The Prompt is what the user inputs, the Generated AtomScript is what the GPT-3 system generates, and the Reference AtomScript is the correct AtomScript to compare to.}
\end{table*}

\appendix

\end{document}